\documentclass[manuscript]{aastex}

\slugcomment{Accepted for publication in ApJ}

\begin{document}

\shortauthors{Chiaberge et al.}
\shorttitle{SEARCH FOR HIGH REDSHIFT FR~I RADIO GALAXIES}

\title{Low-power Radio  Galaxies in the  Distant Universe: A  search for
FR~I at $1<z<2$ in the COSMOS field}

\author{Marco Chiaberge\altaffilmark{1,2}, Grant Tremblay\altaffilmark{1,5}, Alessandro Capetti\altaffilmark{3}, F. Duccio Macchetto\altaffilmark{1}, Paolo Tozzi\altaffilmark{4}, W.~B. Sparks\altaffilmark{1}}
\email{chiab@stsci.edu}

\altaffiltext{1}{Space Telescope Science Institute, 3700 San Martin Drive,
Baltimore, MD 21218}
\altaffiltext{2}{INAF - IRA, Via P. Gobetti 101, I-40129 Bologna, Italy} 
\altaffiltext{3}{INAF - Osservatorio Astronomico di Torino, Via Osservatorio 20, I-10025 Pino Torinese, Italy}
\altaffiltext{4}{INAF - Osservatorio Astronomico di Trieste,  Via Tiepolo 11, I-34143 Trieste, Italy}
\altaffiltext{5}{Rochester Institute of Technology, One Lomb Memorial Drive, Rochester, NY 14623 USA}

\begin{abstract}
We present a search for FR~I radio galaxies between $1 < z < 2$ in the
COSMOS field.  In absence  of spectroscopic redshift measurements, the
selection method  is based  on multiple steps  which make use  of both
radio and optical constraints.  The  basic assumptions are that 1) the
break in  radio power  between low-power FR~Is  and the  more powerful
FR~IIs  does not  change with  redshift, and  2) that  the photometric
properties of  the host  galaxies of low  power radio galaxies  in the
distant universe are  similar to those of FR~IIs  in the same redshift
bin,  as is  the  case for  nearby  radio galaxies.   We describe  the
results  of  our  search,  which  yields  37  low-power  radio  galaxy
candidates that are possibly FR~Is.   We show that a large fraction of
these   low-luminosity  radio   galaxies  display   a   compact  radio
morphology,  that  does  not  correspond  to  the  FR~I  morphological
classification.   Furthermore, our  objects are  apparently associated
with galaxies that show clear  signs of interactions, at odds with the
typical  behavior observed  in low-z  FR~I hosts.   The  compact radio
morphology might  imply that we are observing  intrinsically small and
possibly  young objects, that  will eventually  evolve into  the giant
FR~Is we observe in the local universe.  One of the objects appears as
point-like  in HST  images.   This  might belong  to  a population  of
FR~I-QSOs,  which  however would  represent  a  tiny  minority of  the
overall population of  high-z FR~Is.  As for the  local FR~Is, a large
fraction of  our objects  are likely to  be associated with  groups or
clusters,  making  them  ``beacons''  for high  redshift  clusters  of
galaxies. Our  search for  candidate high-z FR~Is  we present  in this
paper constitutes a pilot study for objects to be observed with future
high-resolution and high-sensitivity instruments  such as the EVLA and
ALMA in  the radio band, HST/WFC3 in  the optical and IR,  JWST in the
IR, as well as future generation X-ray satellites.

\end{abstract}


\section{Introduction}

Among the most energetic phenomena in the universe, radio galaxies are
excellent  laboratories in  which  we investigate  some  of the  major
challenges   of   today's  astrophysics,   such   as  accretion   onto
supermassive   black  holes  (SMBH),   the  associated   formation  of
relativistic    jets   \citep[e.g.][]{blandfordsaasfee,livio99},   the
feedback processes of an ``active'' SMBH in the star formation history
of  a galaxy  \citep[e.g.][]{hopkins06} and  the  role of  the AGN  in
injecting  energy in  the intracluster  medium  \citep{fabian06}.  The
original  classification of  radio galaxies  is based  on  their radio
morphology: ``edge-darkened''  (FR~I), are those in  which the surface
brightness decreases  from the  core to the  edges of the  source, and
typically display large  lobes or plumes; ``edge-brightened'' (FR~II),
are those  in which the  peaks of the  brightness is located  near the
edges of the radio source.  FR~I galaxies typically have a radio power
lower than  that of  FR~II sources, with  the FR~I/FR~II break  set at
$L_{178\mathrm{MHz}}  \sim 2  \times 10^{33}$  erg  s$^{-1}$ Hz$^{-1}$
\citep{fanaroffriley}.  However,  the transition is  rather smooth and
both  radio morphologies  are  present in  the  population of  sources
around  the  break.  The  FR~I/FR~II  break  (at  low redshifts)  also
depends  on   the  luminosity  of   the  host  galaxy,  as   shown  by
\citet{owenledlow94,ledlowowen96}.   However,   it  is  still  unclear
whether  or not that  might simply  be a  result of  selection effects
\citep{scarpaurry01}.   From  the optical  point  of  view, FR~Is  are
invariably  associated with  the most  massive galaxies  in  the local
universe \citep[e.g.][]{zirbel96,donzelli07}, thus  they are also most
likely to  be linked with  the most massive  black holes in  the local
universe.   Furthermore, FR~Is are  usually located  at the  center of
massive  clusters \citep[see e.g.][for  a review]{owenrev96}.   On the
other hand, at low redshifts, FR~IIs are generally found in regions of
lower density, while some FR~II reside in richer groups or clusters at
redshifts higher than $\sim 0.5$ \citep{zirbel97}.

Finding high-$z$ FR~Is and  understanding their evolution will help us
to   address  a  number   of  other   unsolved  problems   in  current
astrophysics,  such  as  studying  the properties  of  the  ``building
blocks'' of  massive elliptical galaxies present  in today's universe,
assessing the relationship between  giant elliptical and their central
supermassive black holes, and  studying the formation and evolution of
galaxy clusters.

However, flux-limited  samples of radio  galaxies such as the  3CR and
its  deeper successors 6C  and 7C  catalogs are  limited by  the tight
redshift-luminosity correlation,  i.e. the well  known Malmquist bias.
This, along with the steep luminosity function of these objects, gives
rise to  a selection  bias resulting in  detection of  high luminosity
sources only  at high redshifts  and low power sources  exclusively at
low  redshift.   It  is  therefore  unsurprising that,  in  the  above
mentioned  catalogs, all  ``high  z'' objects  are  FR~II sources  (or
QSOs),  while  FR~Is  are  only  found at  $z<0.2$.   Indeed,  besides
possibly one  of the two candidates  discussed in \citet{snellenbest},
no FR~I  radio galaxies  are known at  $z>1$.  Nevertheless,  there is
evidence that the population of radio-loud AGN substantially increases
with redshift  up to  $z\sim 2-2.5$ \citep[e.g.][]{ueda03}.   Thus, if
FR~I  galaxies  do in  fact  exist at  high  redshift,  they might  be
significantly  more abundant  at  $z>1$ than  in  the local  universe.
Interestingly,  \citet{sadler07}  find  that  in  the  redshift  range
$0<z<0.7$  radio galaxies  with  radio powers  $P_{1.4}  < 10^{25}$  W
Hz$^{-1}$ undergo  significant evolution.  Their  result is consistent
with being pure  luminosity evolution, which follows a  law similar to
that followed by star forming  galaxies over a similar redshift range.
Clearly, finding  radio galaxies  of low power  at higher  redshift is
extremely  important  to  achieve   a  broader  understanding  of  the
cosmological evolution of these  sources, as compared to the evolution
of normal galaxies.

The  role of  FR~Is in  the framework  of the  unification  scheme for
radio-loud AGN is a significant  matter of debate.  In particular, the
lack of low redshift ``FR~I  quasars'' (defined as radio galaxies with
FR~I morphology  associated with objects showing  broad emission lines
in their  optical-UV spectrum), with  the possible exception of  a few
peculiar    objects,   such   as    the   broad-lined    FR~I   3C~120
\citep[e.g.][]{eracleous94,garcialorenzo05},    is    still   to    be
understood.  It is possible that most, if not all, FR~Is intrinsically
lack a broad  line region and are possibly  powered by low radiatively
efficient  accretion  disks  \citep[e.g.][]{baum95,falcke04,fabian06}.
This  picture  is also  supported  by  the  discovery that  the  large
majority of  FR~I hosts  have faint unresolved  nuclei in  HST images,
whose flux  and luminosity show a  tight correlation with  that of the
radio  cores \citep{pap1}.   The existence  of such  a  correlation is
explained  in  terms  of  a  single  emission  mechanism  (non-thermal
synchrotron radiation  produced at the  base of the  relativistic jet)
for both  the radio and the  optical band \citep{pap1}.   On the other
hand,  recent  work \citep{heywood07}  claims  that  FR~I quasars  are
prevalent in the  universe, based on the analysis of  a sample of QSOs
in  the redshift range  $0.36< z<  2.5$ selected  from the  7C survey,
using  both  their low-frequency  flux  density  and optical  spectral
properties.  However,  some of their  results are still  unclear.  For
example,   these  authors   show   evidence  for   the  existence   of
``high-power''  FR~I  QSO,  whose  nature  has yet  to  be  completely
understood.  A  search for ``bona  fide'' low power radio  galaxies at
$1<z<2$  can clearly  help to  achieve a  better understanding  of the
FR~I-QSO  phenomenon  and  its  role  in  the  framework  of  the  AGN
unification models.

From  the point  of view  of the  environment, FR~I  radiogalaxies are
found in giant ellipticals often  located at the center of clusters of
galaxies.  Finding high-z FR~I  with properties similar to those found
in  the  local universe  can  be a  breakthrough  for  studies of  the
evolution of galaxies and clusters. Using radio galaxies as beacons of
high-z clusters is  not a new idea.  In the  recent past, high-z radio
galaxies  have  often been  used  to  find  protoclusters and  massive
galaxies      at      the      epoch      of      their      formation
\citep[e.g.][]{pentericci01,zirm05,miley08}.   However, all  the above
studies used high power sources with extremely high redshifts ($z>2$).
These are  rare objects  in the universe  whose connection  to today's
radio  galaxies  is not  clear.   It  is  also unclear  whether  their
protocluster environment  have virialized by  that epoch, since  it is
difficult to  detect the X-ray  emission of the ICM.   Powerful FR~IIs
have the disadvantage  of having strong emission from  the nucleus and
powerful relativistic  plasma jets,  which may strongly  influence the
properties  of  the  host  galaxy   and  may  hamper  studies  of  the
environment,     in      particular     in     the      X-ray     band
\citep[e.g][]{fabianxclu03}.  FR~Is  are less powerful  AGNs, they are
more similar to ``normal'' inactive galaxies than FR~IIs, and allow us
an easier investigation of  the surrounding environment, with dramatic
impact on cosmological studies. FR~Is with distorted morphologies were
used  to search  for clusters,  but only  out to  a redshift  $z  < 1$
\citep{blanton00}.   To date,  only  a handful  (less  than 10)  X-ray
confirmed  clusters are  known at  $z>1$,  and none  of them  is at  a
redshift  higher than 1.45  \citep[see][for a  review]{rosati02}.  The
clusters  associated with  our FR~I  candidates might  in fact  be the
``missing link''  between the protoclusters at redshifts  $>2$ and the
well studied clusters of galaxies at $z<1$.

In order  to perform our search  for FR~Is in the  unexplored range of
redshift between z=1 and 2,  we take advantage of the large collection
of  multi-wavelength   data  collected  by   the  COSMOS  collaboration
\citep{scoville07}. In section \ref{overview}  we give an overview of
our method, while in section \ref{selection} we describe our selection
procedure in details.  In section  \ref{irid} we give details about a
few peculiar objects for  which the optical counterpart identification
is uncertain,  but the host galaxy is  clearly seen in the  IR, and in
section \ref{results} we discuss our results from the point of view of
their radio  morphology, local environment  and we show a  few cluster
candidates.   In section \ref{conclusions}  we give  a summary  of the
work, draw conclusions and  outline some future perspectives.

Throughout  this paper  we  use WMAP  cosmology (H$_0$=71,  $\Omega_M$
=0.27, $\Omega_{vac}$  = 0.73). For  the magnitudes of the  sources we
adopt the Vega system.

\section{Overview of our search for high-$z$ FR~Is}
\label{overview}

The  search  for  FR~I  radio  galaxy candidates  between  $1<z<2$  is
performed  using  selection  criteria  in  multiple  wavelengths.   As
already pointed out  in the previous section, flux  limited samples of
radio galaxies include low power sources at low redshift only.  The 3C
sample, which is  arguably the best studied sample  of radio galaxies,
with  his flux  limit set  at  10 Jy  at 178MHz,  includes FR~I  radio
galaxies  only up  to $z  \sim 0.2$.   Deeper catalogs  may  include a
larger number of  distant FR~Is, however, a search  for FR~Is based on
the radio flux only, is extremely inefficient, because of the dominant
population of faint radio  sources associated to e.g. nearby starburst
galaxies.  In other words, by using deep flux limited samples we would
find a multitude of sources whose  radio flux is typical of that of an
FR~I at  $z \sim 1$,  for example, but  are in fact  low-$z$ starburst
galaxies (or, alternatively,  they might even be high  power FR~IIs at
$z > 2$), and very few {\it  bone fide} FR~Is.  In order to select the
right objects,  it is therefore  crucial that our  search discriminate
each candidate  not only  by radio power,  but also by  its properties
across multiple bands.

We focus  our search within the  overlapping fields of  the Very Large
Array Faint Images of the  Radio Sky at Twenty-centimeters (VLA FIRST)
survey \citep{becker95} and the cross-spectrum Cosmic Evolution Survey
(COSMOS,  \citealt{scoville07}).   The  COSMOS field,  a  $1.4^{\circ}
\times    1.4^{\circ}$   square    centered    at   R.A.$=$10:00:28.6,
DEC=+02:12:21.0  (J2000) is  entirely  covered by  FIRST.  The  COSMOS
region  was  selected  because  its  equatorial  position  allows  for
observations from  northern and southern-hemisphere  observatories, as
well as  for its low  and uniform galactic extinction  $\left( \langle
E_{\left(B-V\right)} \rangle \simeq 0.02~\mathrm{mag}\right)$ and lack
of very  bright radio,  UV, and X-ray  sources.  This 2  square degree
region of the sky has been extensively imaged across the spectrum with
deep  observations from  most  of the  major  space- and  ground-based
observatories, yielding a  rich dataset of over 2  million galaxies in
multiple bands.   The specific COSMOS  datasets used in  our selection
procedure   consist  of   1.4   GHz  radio   imaging   from  the   VLA
\citep{schinnerer07}, as well as optical images taken with {\it HST}'s
Advanced Camera  for Surveys (ACS/WFC,  \citealt{koekemoer07}) and the
F814W filter  (similar to  i-band), in $K_s$-band  from the  Kitt Peak
National Observatory (KPNO,  \citealt{capak07}), in optical bands from
Subaru  \citep{taniguchi07},   and  in  $i$  and   $u$-band  from  the
Canada-France-Hawaii  Telescope  (CFHT,  \citealt{capak07}).  We  also
make use of less sensitive  imaging across multiple optical bands from
the  {\it  Sloan Digital  Sky  Survey}  (SDSS, \citealt{york00})  Data
Release 5 (DR5).

Our search procedure begins with the  FIRST survey at 1.4 GHz. The low
resolution FIRST data allow us to easily start our selection procedure
based on the ``total'' radio flux.  FIRST was initially conceived as a
radio sky counterpart to the  Palomar Observatory Sky Survey (POSS I),
and so  encompasses over 10,000  square degrees of the  North Galactic
Cap  (which  includes the  full  COSMOS  field),  imaged in  3  minute
snapshots with $2 \times 7$ 3-MHz frequency channels centered at 1.365
and   1.435  GHz  in   the  VLA's   B-configuration  \citep{becker95}.
Post-pipeline  radio maps  have a  resolution of  $5\arcsec$,  and the
detection threshold  of the  survey is  of order $\sim  1$ mJy  with a
typical RMS  of 0.15 mJy.  At redshift  $z=1.5$ $5\arcsec$ corresponds
to 40 kpc.  Since FR~I radio  galaxies exhibit jet structures at a few
times  100  kpc  scales,  very  little  morphological  information  is
discernible from the 1.4  GHz FIRST images. However, higher resolution
maps  would have  the disadvantage  of missing  some of  the extended,
lower surface  brightness regions, therefore  the FIRST survey  is the
right catalog to begin our search with.

At the 1 mJy detection threshold  of the survey, $\sim 90$ sources per
square degree are  detected.  The low angular resolution  of FIRST is
compensated for  by the VLA-COSMOS  survey \citep{schinnerer07}, which
has a  resolution of $1\arcsec.5 \times  1\arcsec.4$ which corresponds
to $\sim 12$ kpc for a source at $z=1.5$, and a detection threshold of
10  $\mu$Jy (1-sigma).   We therefore  use  the FIRST  images only  to
obtain the  total radio  flux of  the sources. We  use the  deeper and
higher resolution VLA-COSMOS data  to study the actual radio structure
of the sources  that meet our initial flux  selection and derive their
position, as detailed in the next section.

Note  that sources  are not  selected  based on  their ``FR~I''  radio
morphology, but just  using the fact that FR~I  radio galaxies are low
power  radio   sources.  However, we  do use  the radio  morphology to
exclude FR~IIs.   Therefore, at the  end of our selection  process, we
are  left with  a sample  of  low power  radio sources  that are  FR~I
candidates, and whose  radio morphology still has to  be determined in
greater detail.

\section{Selection Procedure}
\label{selection}

\begin{figure*}
\plotone{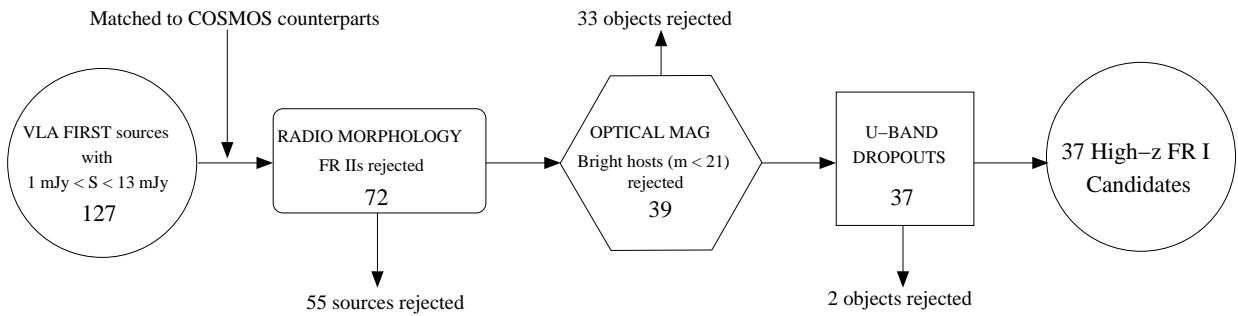}
\caption{Flow-chart describing our  selection procedure. The number of
sources  that survive  each  rejection step  is  reported inside  each
box. See text for more details.}
\label{flowchart}
\end{figure*}

Our  search depends  upon  the assumption  that  the FR~I/FR~II  break
luminosity  at  $L_{1.4~\mathrm{GHz}}   \sim  4  \times  10^{32}$  erg
s$^{-1}$ Hz$^{-1}$ (assuming a  spectral index $\alpha_r =0.8$ between
1.4  GHz and 178  MHz) does  not change  with redshift.   Moreover, we
assume that the photometric  properties of high-$z$ FR~I host galaxies
are similar  to those of  FR~II hosts in  the same range  of redshift.
Note  that  photometric  redshifts  for  COSMOS  sources  are  already
available in the literature \citet{mobasher07}.  Photometric redshifts
have proved  to be  statistically reliable for  a large  population of
objects,  but for  single  objects  (and for  AGN  in particular)  the
uncertainty of $z_{phot}$ can be large.  Therefore, we use photometric
redshifts only as a  confidence check (see Sect.\ref{results}), and not
to  confirm or  discriminate among  candidates.  That  is,  no objects
meeting  our  selection criteria  are  rejected  based on  photometric
redshift value.  Based on  these two basic assumptions outlined above,
we  describe our selection  procedure in  the following.  A flow-chart
describing our selection procedure is shown in Fig~\ref{flowchart}.

1. {\bf Radio  selection: flux limits.} We select  FIRST radio sources
inside the COSMOS  field whose 1.4 GHz total  flux corresponds to that
expected  for FR~Is  at $1<z<2$.   To that  end, we  require  that our
candidate sources reside in a  narrow bin of $L_{1.4}$.  The limits of
this ``allowable''  1.4 GHz luminosity  range ($L_{\mathrm{1.4,min}}$,
$L_{\mathrm{1.4,max}}$) are set such that the objects we select have a
total radio power typical of FR~Is.   That is, the radio power must be
significantly  {\it above}  the level  of radio  activity  produced by
``non  radio-loud   AGN''  and  starburst  galaxies\footnote{Starburst
  galaxies typically  have $L_{1.4} < 10^{30}$  erg s$^{-1}$ Hz$^{-1}$.}
in  the target  redshift bin,  but safely  {\it below}  the FR~I/FR~II
break.  Based  on these criteria,  we calculate $L_{\mathrm{1.4,min}}$
and    $L_{\mathrm{1.4,max}}$     in    terms    of     flux    limits
$F_{\mathrm{1.4,min}}$    and    $F_{\mathrm{1.4,max}}$.     We    set
$F_{\mathrm{1.4,min}}$  to  correspond to  the  flux  observed from  a
source  of  radio power  $L_{\mathrm{1.4,min}}$  at  $z=2$ (e.g.   the
faintest   of   the   objects    we're   searching   for),   and   set
$F_{\mathrm{1.4,max}}$  to   the  flux  of  a   source  of  brightness
$L_{\mathrm{1.4,max}}$ at  $z=1$ (e.g.  the very  brightest sources we
wish  to   find).   The  observed  flux   of  the  $\nu   =  1.4$  GHz
Fanaroff-Riley break luminosity  for each value of $z$  can be derived
using
\begin{equation}
F_{\nu}\left(\nu_{0}\right)=\frac{L_{\nu}\left(\nu_{0}\right)\left(1+z\right)^{1-\alpha}}{4 \pi D^{2}_{L}},
\end{equation}
where $L_{\nu}(\nu_{0})$ is the  luminosity of the FR~I/FR~II break at
$\nu_{0}$.  We assume a spectral index $\alpha=0.8$. In the formula, a
factor of $\left(1+z\right)$ accounts  for the passband and another of
$\left(1+z\right)^{-\alpha}$ makes up the $K$ correction.  The flux of
the FR~I/FR~II break  at $z=1$ is $F_{1.4} \sim  26$ mJy.  However, we
set our upper ``allowable  flux'' limit $F_{\mathrm{max}}$ a factor of
two fainter than  this (to 13 mJy)  to ensure that, at the  end of the
process,   we  select   ``bona  fide''   FR~I  candidates   and  avoid
accidentally  selecting  FR~IIs near  the  break luminosity.  However,
these  FR~IIs  are  rejected  during  step 2.  based  on  their  radio
morphology. The flux of the break at $z=2$ is $\sim 10$ mJy, and so we
set our  lower limit  $F_{\mathrm{min}}$ to be  an order  of magnitude
fainter (at 1 mJy). This corresponds to the detection threshold of the
FIRST     survey.      We      searched     the     FIRST     database
(\texttt{http://sundog.stsci.edu/})   for  radio   sources  possessing
integrated 1.4 GHz  fluxes between 1 mJy and 13  mJy within the COSMOS
field.   The  number  of  sources  that match  our  flux  criteria  is
131. Clearly, a  selection based on  flux only allows the  presence of
both $z<1$ ``faint'' sources  and $z>2$ ``powerful'' sources. The next
selection steps are  designed to reject most of  the objects that fall
outside our preferred redshift range.

\begin{figure*}
\epsscale{0.8}
\plotone{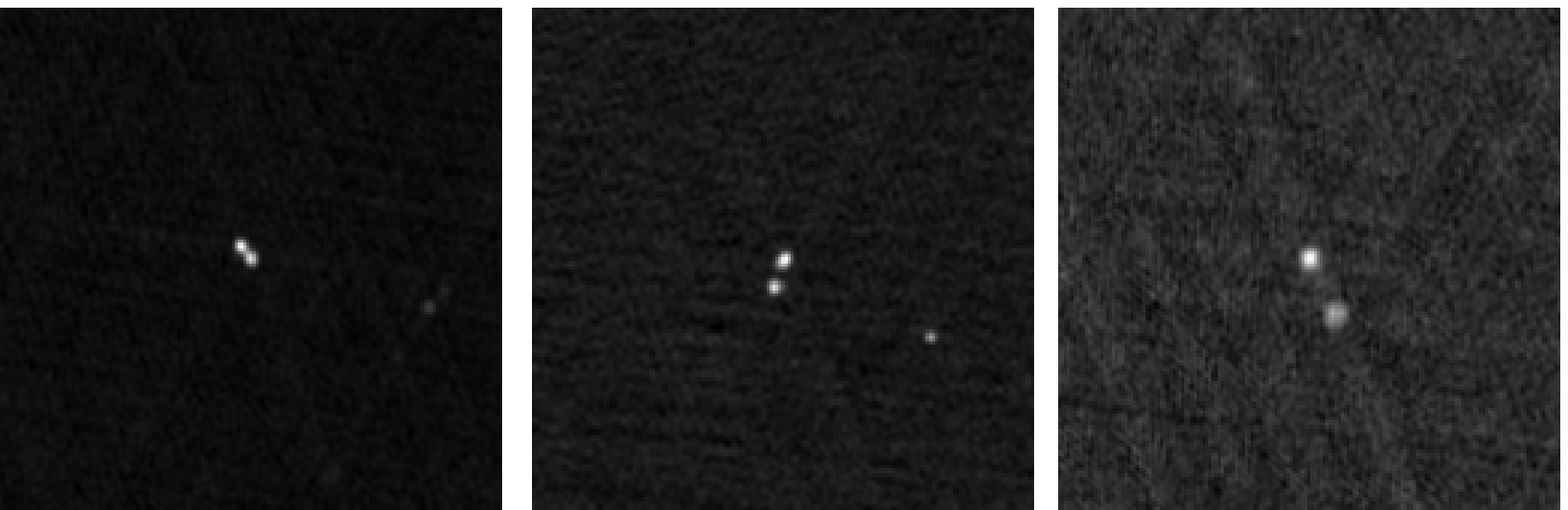}
\caption{Examples of sources showing a clear double (FR~II) morphology
that were  rejected during  our selection step  2. The objects  are the
FIRST   sources   J095908+024809   (left),   J100217+012220   (center)
J100245+024534 (right). The  size of each image is  5'x5', as obtained
from      the      FIRST       image      cutouts      archive      at
\texttt{http://third.ucllnl.org/cgi-bin/firstcutout}.}
\label{frII}
\end{figure*}

\indent 2.   {\bf Radio selection: morphology.}   The sources selected
at step 1 are  individually examined for large-scale radio morphology.
Those   candidates   featuring   clearly   ``edge-brightened''   radio
structures  are rejected,  so as  to  filter out  more powerful  FR~II
sources  at  redshift  $z>2$   which  have  passed  our  initial  flux
selection.  Though  the FIRST  imaging is at  very low  resolution, it
typically  suffices for  the identification  of classical  doubles, as
edge-brightened lobes  or hotspots of  FR~IIs are found on  $>$100 kpc
scales, which translates into $>11^{\arcsec}$ at $z=2$. As a result of
this, a significant number of pairs of radio sources that were counted
as two ``single''  objects at the previous step  are now recognized to
be ``double''  sources. Twenty-two of those FR~IIs  are identified. In
Fig.~\ref{frII}  we show three  examples of  such sources.   Note that
these account  for 44 ``single''  FIRST radio sources.  We  then check
the VLA-COSMOS  radio maps  to make sure  that objects that  appear as
compact in  the FIRST  images are not  smaller ``double''  FR~II radio
sources. Eleven FR~IIs are identified using VLA-COSMOS radio maps, for
a total of 55 such objects rejected based on their radio morphology.

\indent  3.  {\bf  Optical  selection: magnitude.}   In  order to  set
constraints  on the  host galaxies'  photometric properties,  the next
step involves  the identification of  the optical counterparts  of the
radio  sources.  Therefore,  for our  sample, the  optical counterpart
identification is part of our selection process.  The method we use is
simply to blink the COSMOS-VLA  radio data with the optical COSMOS-HST
image,  registered  on the  same  WCS  framework.   Despite the  short
exposure time  (single orbit observations), the  COSMOS-HST images are
the most suitable data  for identifying the optical counterparts.  The
significantly  higher angular  resolution, as  compared to  the ground
based optical data,  allows us to avoid confusion.   In most cases, it
is straightforward to identify the  host galaxy, since the position of
the radio core is well set by the COSMOS-VLA images, and the beam size
is small enough that only one  galaxy is found at the same position on
the   HST   image.   We will  discuss a  few peculiar  cases for  which the
optical counterparts  are not  easily detected in  section \ref{irid}.
We  also unambiguously  identify the  host galaxy  for 24  FR~IIs.  We
check the  COSMOS source catalog  at the coordinates  corresponding to
the  radio sources that  match the  first two  selection steps  and we
obtain the magnitudes of each object in different bands.

\begin{figure}
\epsscale{0.8}
\plotone{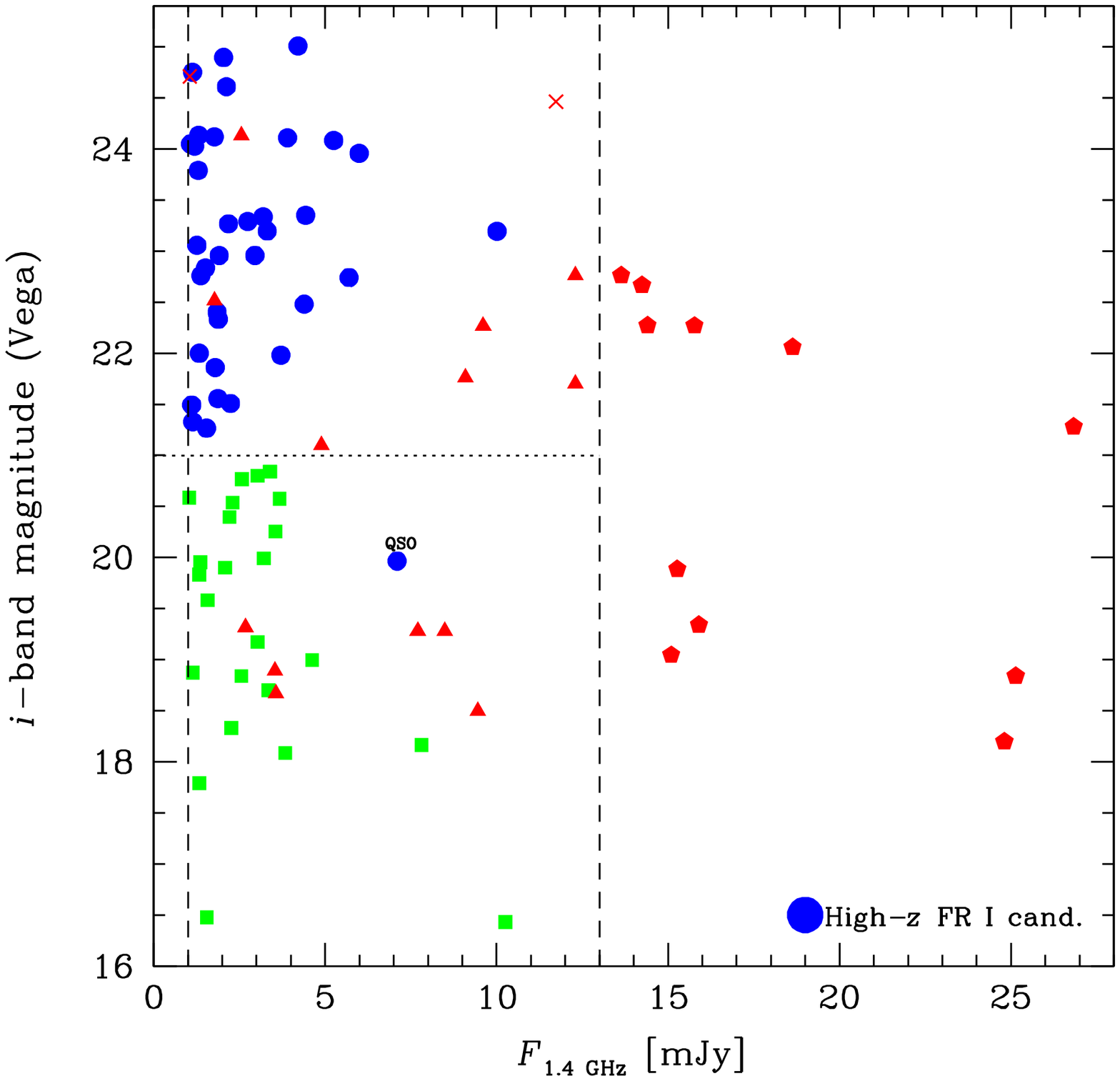}
\caption{Optical i-band  magnitude plotted  against the radio  flux at
  1.4  GHz.  Only the  sources  for which  a  host  galaxy is  clearly
  detected are  plotted. The two  vertical dashed lines are  the radio
  flux limits  for the  selection process at  step 1.   The horizontal
  line  represents our  lower limit  for the  optical  selection (step
  3). Circles are the  FR~I candidates, triangles are rejected because
  of  their FR~II radio  morphology, pentagons  are rejected  by radio
  flux, squares are rejected by  host galaxy magnitude and crosses are
  u-band  dropouts.  Note  that the  QSO is  not rejected  despite its
  bright optical magnitude (see text).}
\label{magflux}
\end{figure}

We set  a lower limit in  optical (i-band) magnitude  to reject fairly
bright low  redshift galaxies  with intrinsically less  powerful radio
emission (e.g.  possibly star forming galaxies).  In this step we make
use  of the assumption  that the  properties of  the host  galaxies of
FR~Is  are  similar  to  those  of  FR~IIs, as  it  is  the  case  for
low-redshift radio  galaxies.  The K-band magnitude of  an FR~II radio
galaxy   at  $1<z<2$   is  known   not   to  exceed   $M_K  \sim   17$
\citep[e.g.][]{willott03}, and the typical I-K color of FR~II hosts is
$\sim 4$ or  higher.  This sets a lower limit  to the i-band magnitude
of  $\sim 21$.   In Fig.~\ref{magflux}  we plot  the  i-band magnitude
against the  radio flux for  the sources with an  optical counterpart.
Note that  not all of the FR~IIs  are plotted in the  diagram: some of
the FRIIs are left out because  either the host galaxy is not detected
in the optical,  or it cannot be identified  because it lies somewhere
between the location  of the two radio hotspots  and multiple galaxies
are present  in the same  region. For the  candidate FRIs that  do not
have an optical  counterpart but for which the host  galaxy is seen in
the IR only (see Sect~\ref{irid}), we plot the magnitude of the closest
optical  source. This  can be  interpreted just  as a  lower  limit to
magnitude of the  host and it is completely  irrelevant from the point
of view  of the  sample selection, since  we do not  discard optically
faint objects. 

Note  that, as  a result  of  the magnitude  limit, most  of the  host
galaxies of our  candidates are not detected in  the Sloan Digital Sky
Survey.  This step in the selection process is only intended to reject
those  host galaxies  that  are unreasonably  brighter  or bluer  than
typical  radio galaxies  in the  target redshift  range.  Importantly,
setting  a  limit in  i-band  ensures us  that  we  are not  rejecting
galaxies that  are fainter than  ``normal'' radio galaxies  in K-band.
With this  selection step we  only filter out bright  nearby galaxies,
and  we  are keeping  in  our  sample  distant and  intrinsically  red
objects. Thirty-three  objects are rejected  at this stage  because of
the host galaxy optical brightness.

One object  with $m_I <  21$ (object 236)  is not rejected  because it
appears to be ``stellar-like'' in  the HST image.  Since it completely
lacks the  host galaxy, the magnitude  of the point  source we observe
can be considered as a lower  limit to the magnitude of the host.  The
nature of this source is unclear:  one possibility is that it is a QSO
that resides in the redshift  range $1<z<2$.  In that case, because of
its  low  radio power,  it  would  represent  an interesting  case  of
FR~I-QSO  (see Section  \ref{results}). Alternatively,  it could  be a
high-power radio-loud QSO located at a redshift higher than 2.

\indent 4. {\bf Optical selection:  u-band dropouts.} As the last step
in  the selection  process, we  want to  make sure  to include  in our
sample objects with redshift not  significantly higher than 2. This is
mainly  because the  radio power  of  those objects  would exceed  the
FR~I/FR~II break.  We check the deep COSMOS ground based images and we
reject two sources  that are detected in V and B-band,  but not in the
U-band.  Such ``$u$-band dropouts''  are typically galaxies located at
a redshift  significantly higher  than our range  of interest,  with a
lower limit at $z\sim 2.5$ \citep[e.g.][]{giavalisco02}.\\

At the  end of the selection  process we are left  with 37 candidates,
that  are listed  in Table~\ref{tab1}.   The Table  reports,  for each
source,  the radio  flux  at 1.4  GHz,  the magnitude  of the  optical
counterpart  in  K$_s$,  V  and  i  band (in  the  Vega  system),  the
photometric redshift of the source as it appears in the COSMOS catalog
\citep{mobasher07},  and  a qualitative  description  of the  observed
radio and optical morphologies.

\begin{figure}
\epsscale{0.8}
\plotone{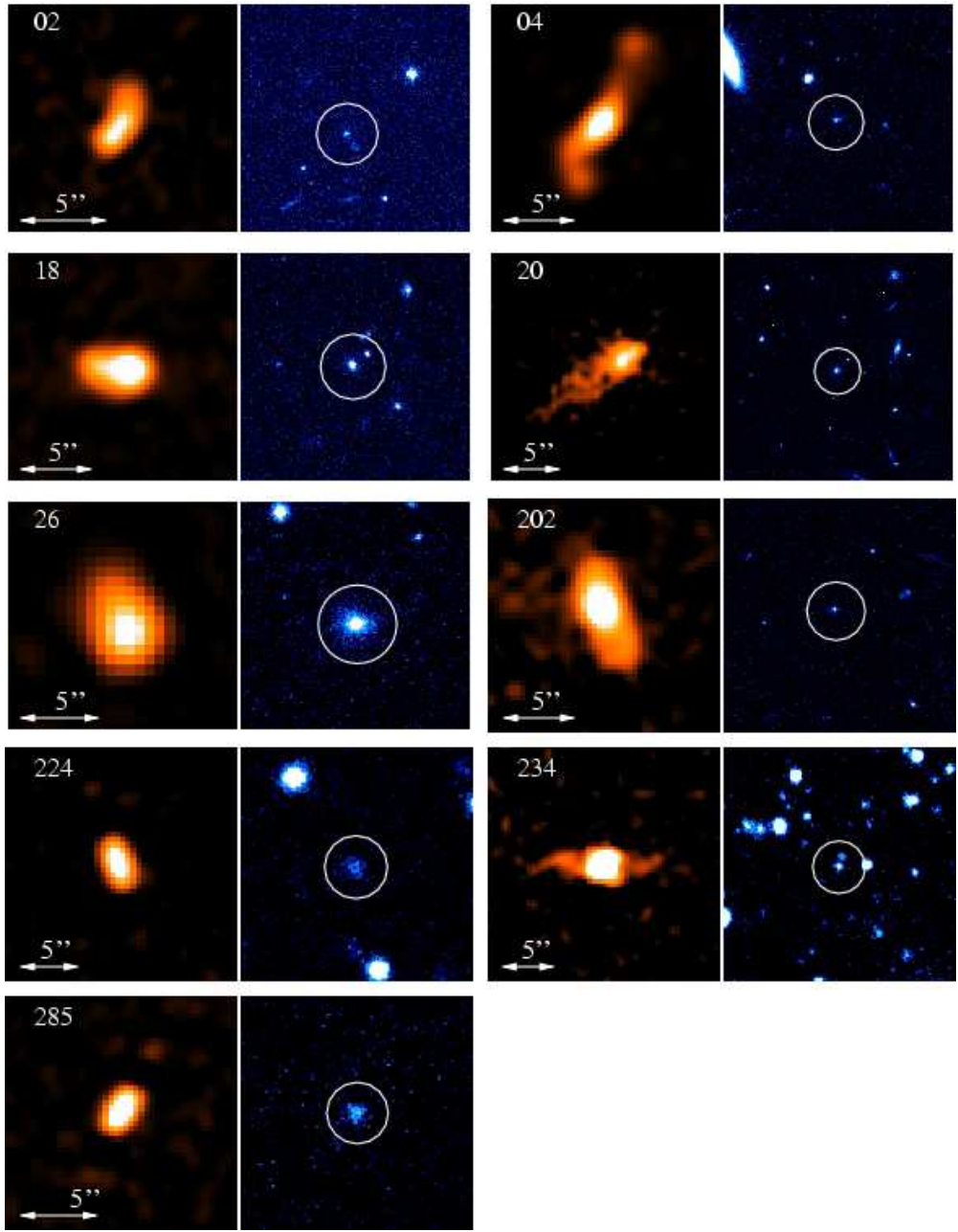}
\caption{FR~I candidates  with extended 1.4 GHz  radio morphology.  In
these images,  5'' corresponds to a  linear scale of $\sim  40$ kpc at
$z=1.5$.   For each  object,  the image  in  the left  panel are  from
VLA-COSMOS survey, while in the right panel we show the HST-COSMOS ACS
images (F814W), except for 234 and 285 where the Subaru i-band
image is shown.}
\label{extended}
\end{figure}

\begin{figure}
\epsscale{0.8}
\plottwo{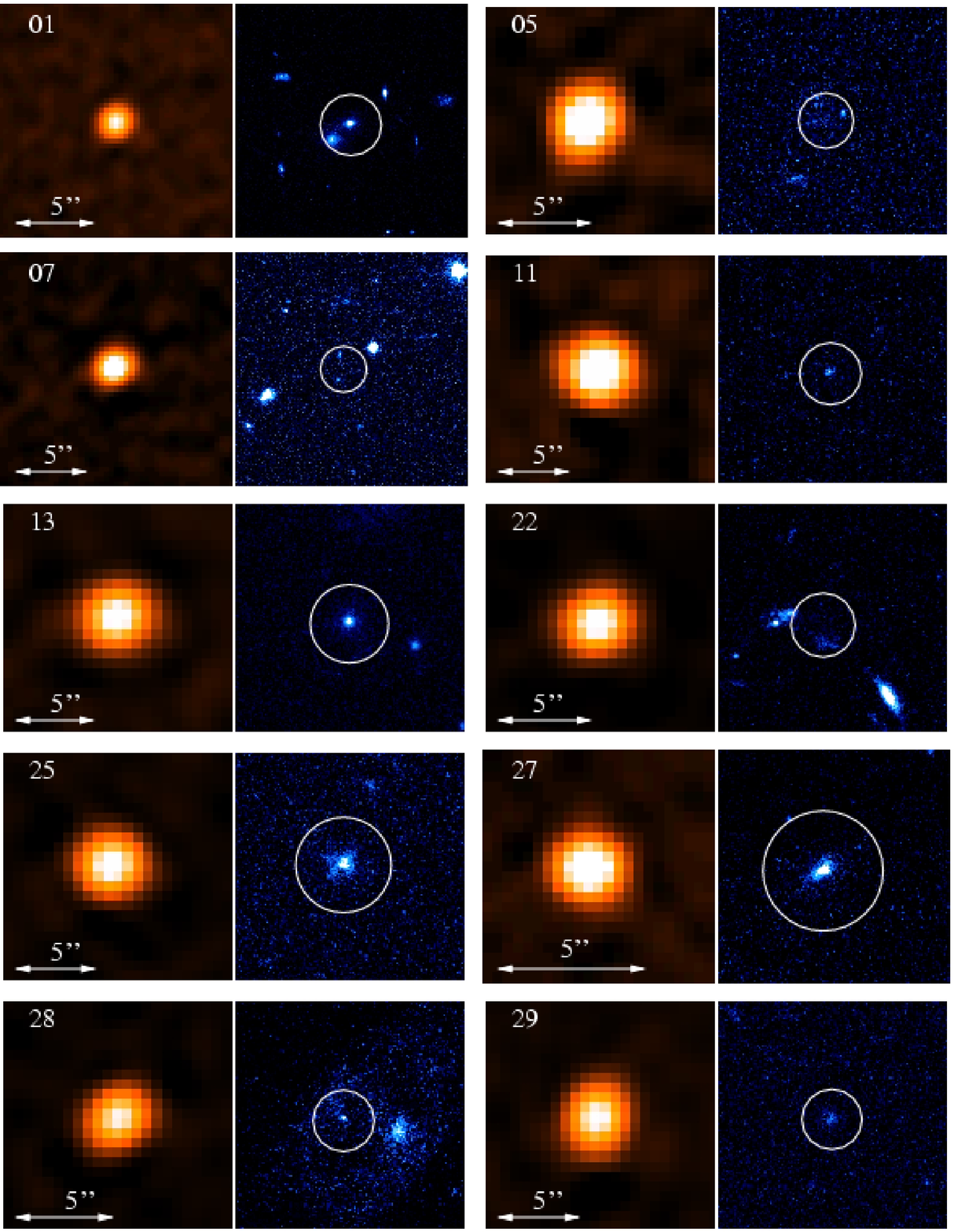}{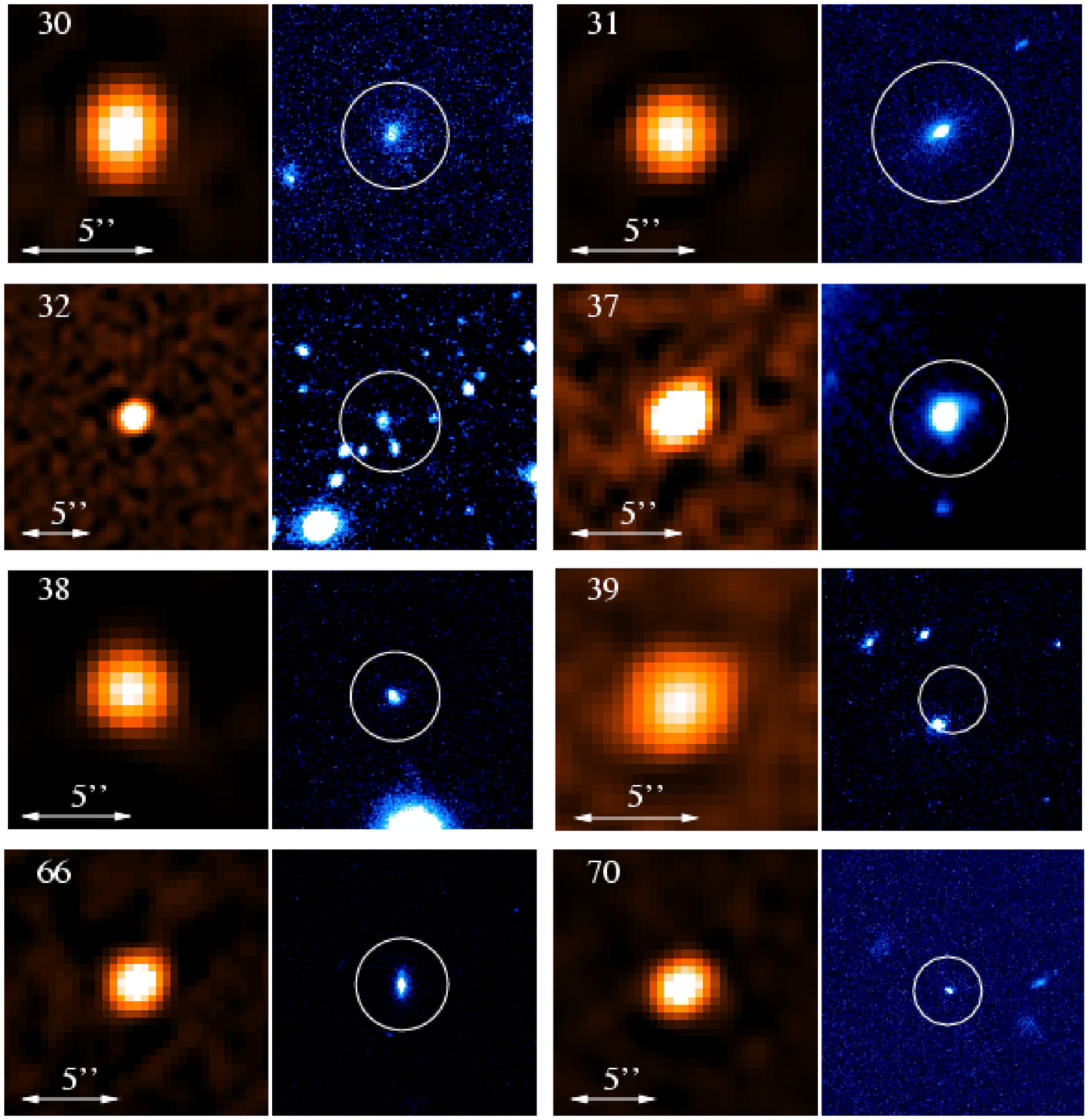}
\plotone{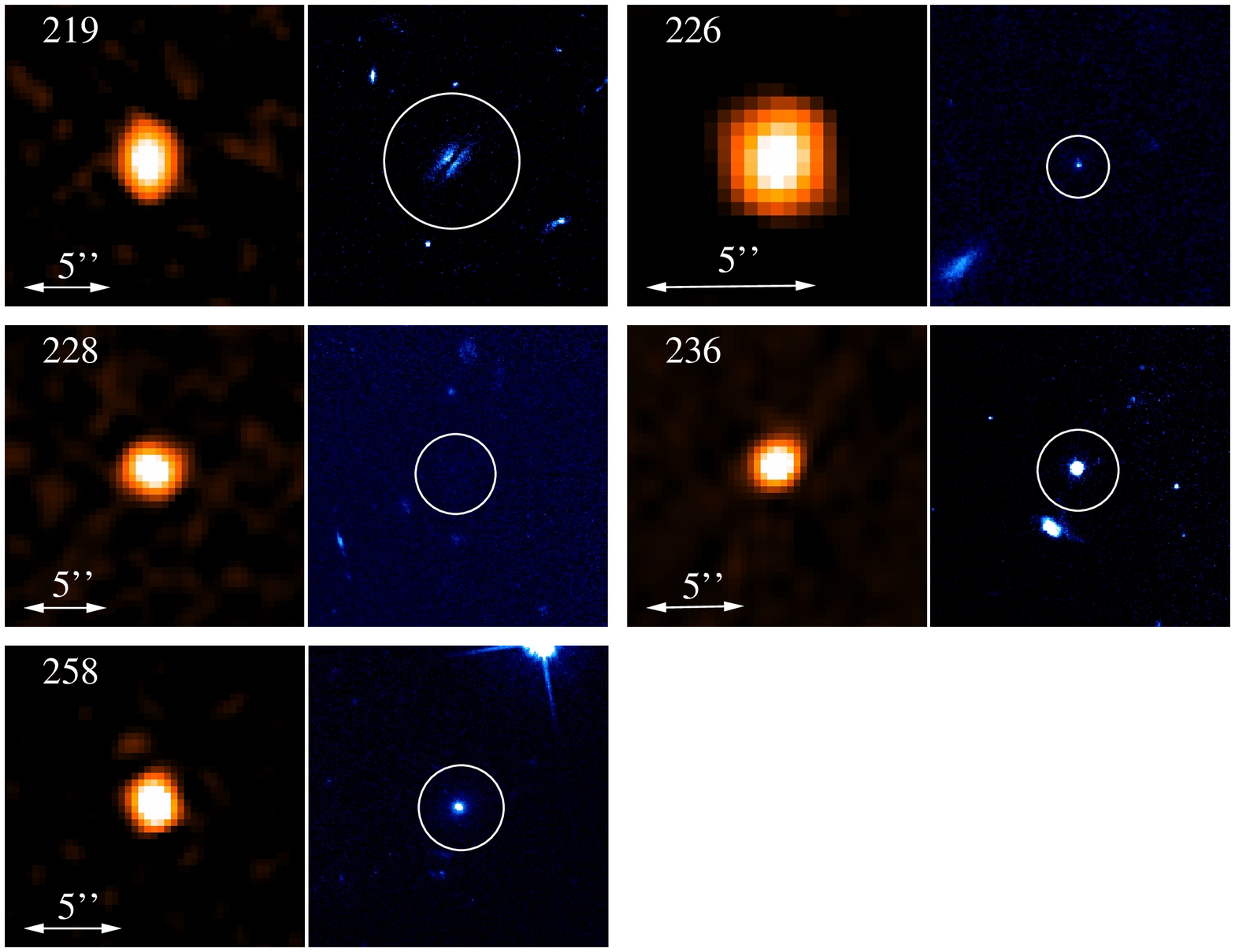}
\caption{FR~I candidates  with compact (only slightly  resolved at the
resolution  provided by  the VLA  images) radio  morphology.  For each
object, the image  in the left panel is  from VLA-COSMOS survey, while
in the right panel we  show the HST-COSMOS ACS images (F814W),
except for objects  32 and 37 where the Subaru  i-band image is shown,
since the HST image is not available for those two objects.}
\label{compact}
\end{figure}

\begin{figure}
\plotone{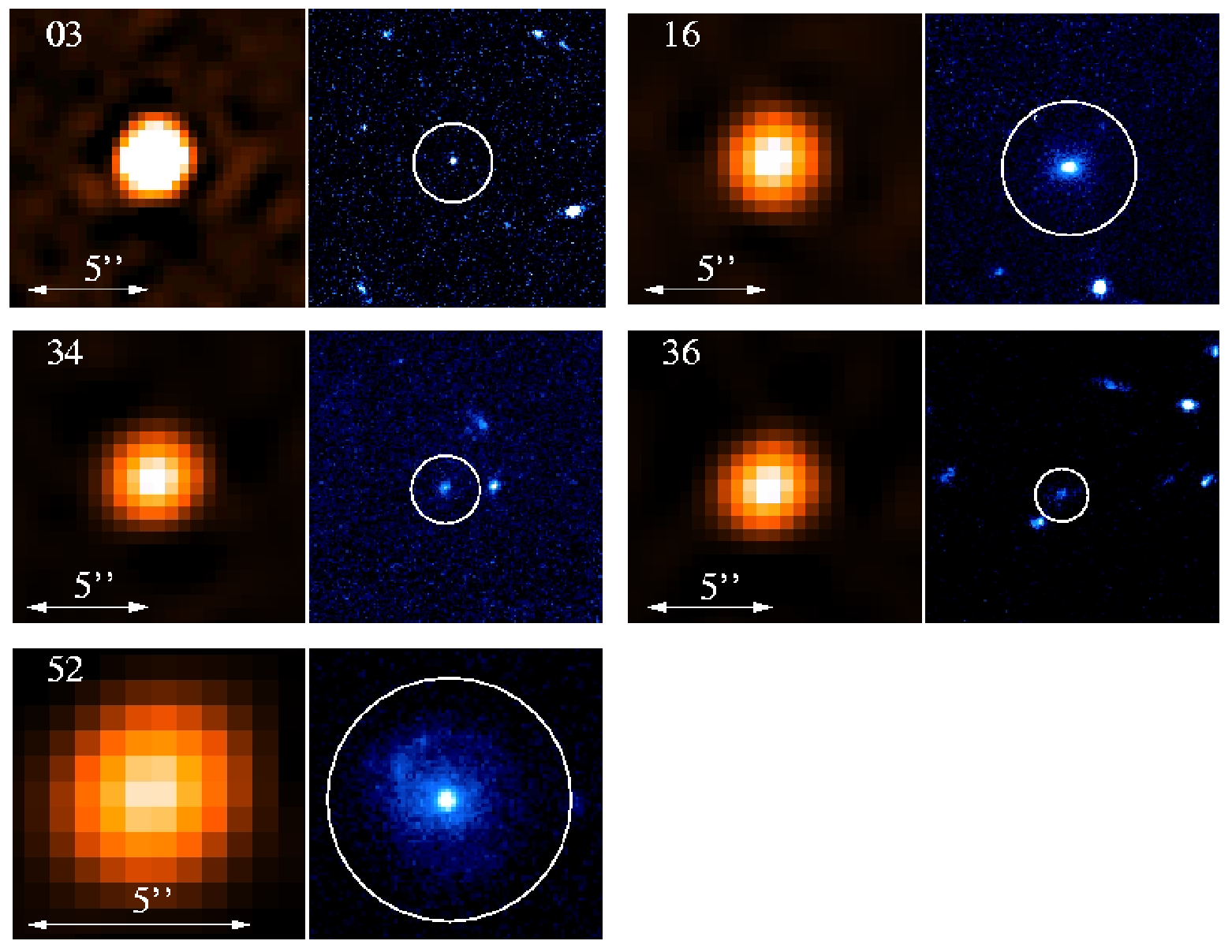}
\caption{FR~I candidates with unresolved radio morphology (the FWHM of
the source  is consistent with the  beam width of the  VLA image). For
each object, the image in the left panel is from VLA-COSMOS survey. In
the right panel we show the HST-COSMOS ACS images (F814W)}
\label{unres}
\end{figure}

Radio and optical images for each  of the FR~I candidates are shown in
Figs.~\ref{extended}, \ref{compact} and \ref{unres}. The radio data at
1.4  GHz are  taken from  the VLA-COSMOS  survey \citep{schinnerer07}.
The  optical  images  are   from  the  HST-COSMOS  programs  (GO~9822,
GO~10092)   and   were  taken   using   {\it   HST}/ACS  in   $i$-band
\citep{koekemoer07}, unless mentioned otherwise.

In  the following  section we  discuss issues  related to  the optical
counterpart  identifications.   Note that  the  identification of  the
optical counterparts is  part of our selection process  at steps 3 and
4. In fact, we  look both at the shallow SDSS  images (mainly to check
that no optical  counterpart is found, as prescribed  by our selection
step  2)  and  at the  deeper  ground  based  images from  the  COSMOS
collaboration  (to identify  the  optical counterpart).   In the  next
section  we describe  a few  peculiar cases  of objects  that  show no
optical counterpart but are clearly detected in th IR.

\section{IR identifications}
\label{irid}

\begin{figure}
\epsscale{0.8}
\plotone{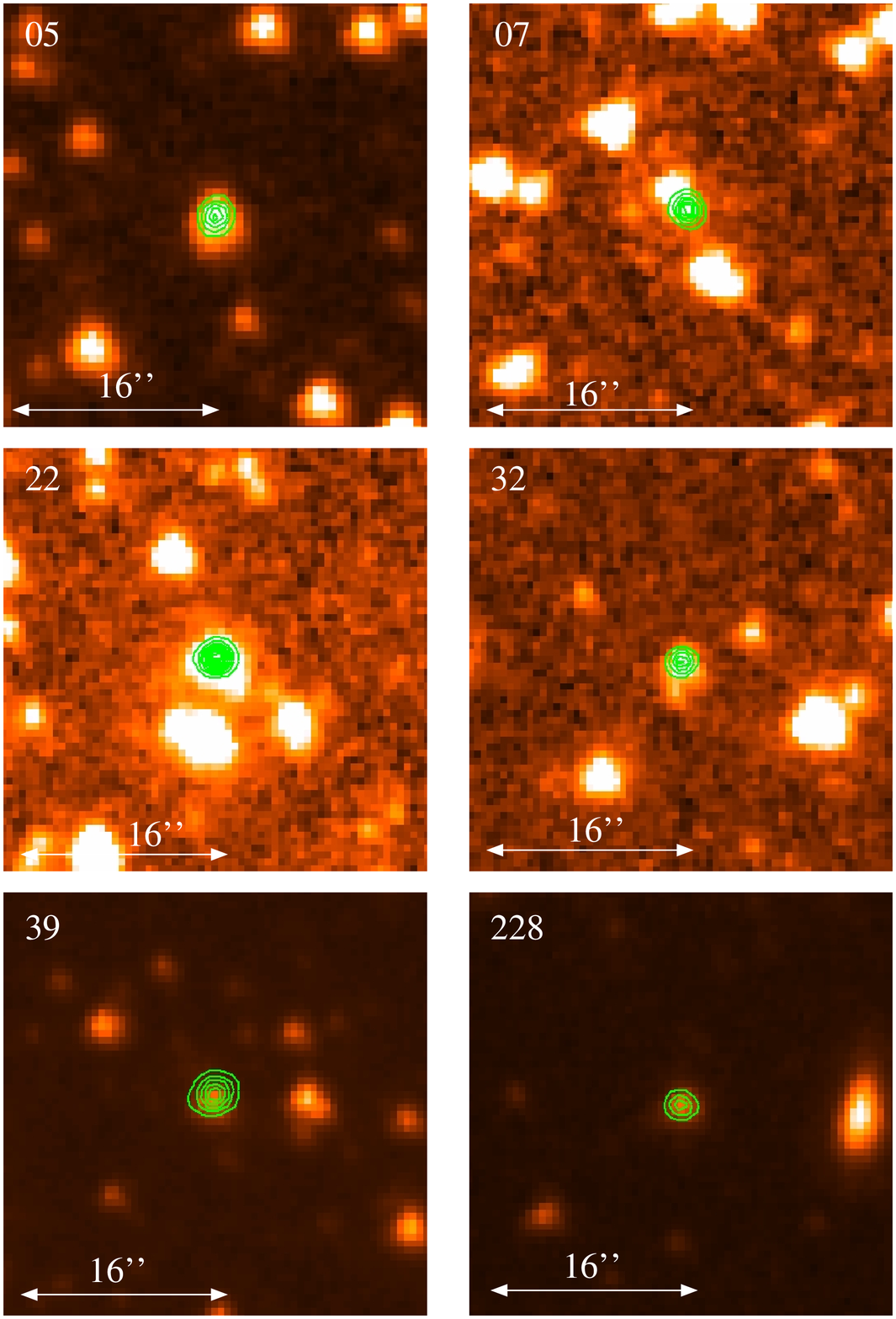}
\caption{Objects identified  using {\it Spitzer} IR  images. The radio
contours are overplotted onto the  3.6$\mu$ image to show the location
of the source.}
\label{spitzer}
\end{figure}

In  a  few cases  the  optical  identifications  are difficult.   This
generally happens because either  no obvious optical source co-spatial
with the radio core is seen, or  it is not clear whether we are seeing
a group  of galaxies  or a single  irregularly shaped object.   In six
cases, the  peak of  the optical emission  is not coincident  with the
peak  of the radio  core. This  is clearly  shown in  the case  of our
candidates 5,  7, 22, 32,  39 and 228 (see  Figs.~\ref{compact}).  For
object 5,  a very  low surface brightness  object is  only tentatively
detected  in the  ACS image,  even after  significant  smoothing.  For
object  22, the optical  counterpart is  also not  clearly identified,
since 3 to 4 relatively bright  galaxies are present in the ACS image,
but none of  them appears to be co-spatial with the  peak of the radio
core.  For  objects 7, 39 and  228 no optical  counterpart is detected
anywhere  near the  location of  the radio  source. For  object  32 an
optical source is  clearly seen in the Subaru i-band  image, but it is
located $\sim 2\arcsec$ E of the radio source.  However, in all cases,
{\it Spitzer Space Telescope}/IRAC images  taken as part of the COSMOS
program  \citep{sanders07}  clearly  reveal  the host  galaxy  at  the
location of the radio source.   In Fig.~\ref{spitzer} we show the {\it
Spitzer} images at  3.6 $\mu$m, together with the  radio contours that
indicate  the location of  the FR~I  candidate.  Clearly,  the optical
magnitude  listed in Table~\ref{tab1}  for the  objects that  are only
identified in the IR should be consider as a lower limit, since it has
been derived from  the COSMOS catalog from the  optical object closest
to the radio source.

The lack of an optical counterpart, together with the detection of the
IR counterpart  is intriguing.  The absence  of the object  in the ACS
image might  be explained by the fact  that the host galaxy  is a very
low surface brightness object, thus the short exposure time of the HST
images, together with the small  collecting area of the telescope does
not allow us  to detect it.  However, we would expect  to detect it in
the deeper  ground based images.  This cannot  be clearly established,
mostly because of  confusion problems.  In the IR  Spitzer images, the
elliptical  host can  be more  easily  seen, since  it dominates  the
emission with respect to the bluer surrounding galaxies that disappear
in the  IR.  An alternate scenario  is that these  are higher redshift
objects,  not visible in  the optical  because of  Hydrogen absorption
(i-band  dropouts).  In  this case,  the  bluer galaxies  seen in  the
optical  data around the  position corresponding  to the  radio source
would not  be at  the same redshift  as the  host galaxy of  the radio
source.  For an i-band  dropout, the  redshift of  the host  would be
around $z\sim  6$, therefore these  objects would be high  power radio
sources  that   should  not  be  included  in   our  sample.   Further
investigation is needed to assess their nature.  It is also worth
noting that  for these objects with  uncertain optical identification,
$z_{phot}$ is most likely derived from one of the galaxies surrounding
the real host.

\begin{deluxetable}{cccccccc}
\tablecolumns{8}
  \tablecaption{$1 < z < 2$ FR I Radio Galaxy Candidates}
  \tablewidth{0pc}
  \tablehead{
    \colhead{ID} &
    \colhead{$K_s$ Mag} &
    \colhead{$V$ Mag} &
    \colhead{$I$ Mag} &
    \colhead{$S_{\mathrm{20~cm}}$ (mJy)} &
    \colhead{$z_{\mathrm{phot}}$} &
    \colhead{Radio Morphology} & 
    \colhead{Optical Morphology}\\
    \colhead{(1)} & \colhead{(2)} & \colhead{(3)} &
    \colhead{(4)} & \colhead{(5)} & \colhead{(6)} &
    \colhead{(7)} & \colhead{(8)} 
  }
  \startdata
01 &17.710   &24.423  &21.860  & 1.79  & 0.94  &Compact & Smooth  \\
02 &19.279   &25.835  &24.047  & 1.08  & 1.59  &Extended & Compact \\
03 &21.014   &25.583  &25.008  & 4.21  & 1.59  &Unresolved & Compact \\
04 &18.872   &25.473  &23.957  & 5.99  & 1.85  &Extended & Compact  \\
05 &18.991   &24.449  &23.789  & 1.30  & 2.08  &Compact &  -- \\
07 &20.072   &24.942  & 23.777 & 1.14  & 1.09  &Compact &  -- \\
11 &19.716   &26.994  & 24.750 & 1.13  & 1.05  &Compact & Compact \\
13 &18.670   &24.711  &22.835  & 1.51  & 1.21  &Compact & Compact \\
16 &18.668   &24.860  &22.741  & 5.70  & 1.10  &Unresolved & Smooth \\
18 &19.015   &24.316  &22.479  & 4.39  & 0.93  &Extended & Complex \\
20 &18.276   &24.594  &21.998  & 1.33  & 0.98  &Extended &Compact \\
22 &19.698   &24.043  &23.288  & 2.74  & 1.51  &Compact & -- \\
25 &18.787   &24.952  &23.266  & 2.18  & 1.40  &Compact & Complex  \\
26 &17.631   &24.908  &22.332  & 1.88  & 1.30  &Extended &Smooth \\
27 &18.722   &24.279  &22.957  & 1.91  & 1.39  &Compact &Complex \\
28 &20.158   &25.127  &24.118  & 1.77  & 1.23  &Compact & Compact \\
29 &21.099   &25.341  &24.610  & 2.12  & 1.03  &Compact &Compact  \\
30 &18.360   &25.812  &23.055  & 1.26  & 1.15  &Compact &Complex \\
31 &18.456   &23.948  &21.981  & 3.71  & 0.88  &Compact &Smooth \\
32 &20.214   &25.095  &24.134  & 1.31  & 2.17  &Compact & Compact \\
34 &19.105   & 25.152 &24.082  & 5.25  & 2.04  &Unresolved &Compact  \\
36 &18.606   & 24.782 &23.335  & 3.19  & 1.42  &Unresolved & Complex  \\
37 &18.176   &22.388  &21.556  & 1.87  & 1.26  &Compact &Smooth \\
38 &19.489   &24.603  &23.193  & 10.01 & 1.15  &Compact & Complex \\
39 &18.405   &25.268  &22.759  & 1.37  & 1.36     & Compact          & -- \\
52 &17.928   &23.132  &21.266  & 1.54  & 0.84      &  Unresolved         & Complex  \\
66 &18.149   &23.637  &21.493  & 1.11  & 0.80      &  Compact         & Smooth \\
70 &  19.521 & 24.766 & 24.109 &  3.90 & 2.75 &  Compact   &  --  \\
202 & 19.706 &26.376  &24.049  &  1.08 & 1.24     &    Extended       & Compact \\
219 &18.256   &24.517  &22.402  & 1.85 & 1.20     &  Compact         &Complex \\
224 &18.636   &25.414  &23.196  & 3.31 & 1.40     &    Extended       &Compact \\
226 &19.879   &25.225  &24.027  & 1.19 & 2.04     & Compact          & Compact \\
228 &19.379   &27.163  &24.894  & 2.04 & 1.45      &  Compact         & -- \\
234 &18.724   &25.399  &23.350  & 4.43 & 1.42     &    Extended       &Complex \\
236 &17.461   &20.594  &19.965  & 7.10 & 1.23     &  Compact         &  QSO \\
258 &17.860   &23.190 &21.508 &   2.24 & 1.07      &  Compact         & Compact \\
285 &19.018   &24.022  &22.958 &  2.95 & 1.24      &  Extended         &Complex \\
  \enddata
  \tablecomments{\tiny{
    (1) Object ID;  
    (2) $K_{s}$-band apparent magnitude in the Vega system;
    (3) $V$-band apparent magnitude (Vega). The magnitude for the objects identified in the IR only are that of the closest optical counterpart;
    (4) $I$-band apparent magnitude (Vega). The magnitude for the objects identified in the IR only are that of the closest optical counterpart \citep{capak07};
    (5) Integrated radio flux at 20 cm (mJy) from the FIRST survey;
    (6) Photometric redshift $z_{\mathrm{phot}}$, calculated by \citet{mobasher07};
    (7) Qualitative characterization of radio morphology, based on VLA-COSMOS image. 
This classification reflects whether the corresponding image of this target is located in 
Figs.~\ref{extended},   \ref{compact},   \ref{unres} ;  
(8) Qualitative characterization of the morphology of the optical counterpart 
to the radio source, based on inspection of the ACS $I$-band image. The optical morphology classification for 
the host galaxies detected only in the IR is omitted. 
 } }
\label{tab1}
\end{deluxetable}

\section{Results and discussion}
\label{results}

\begin{figure}
\epsscale{0.8}
\plotone{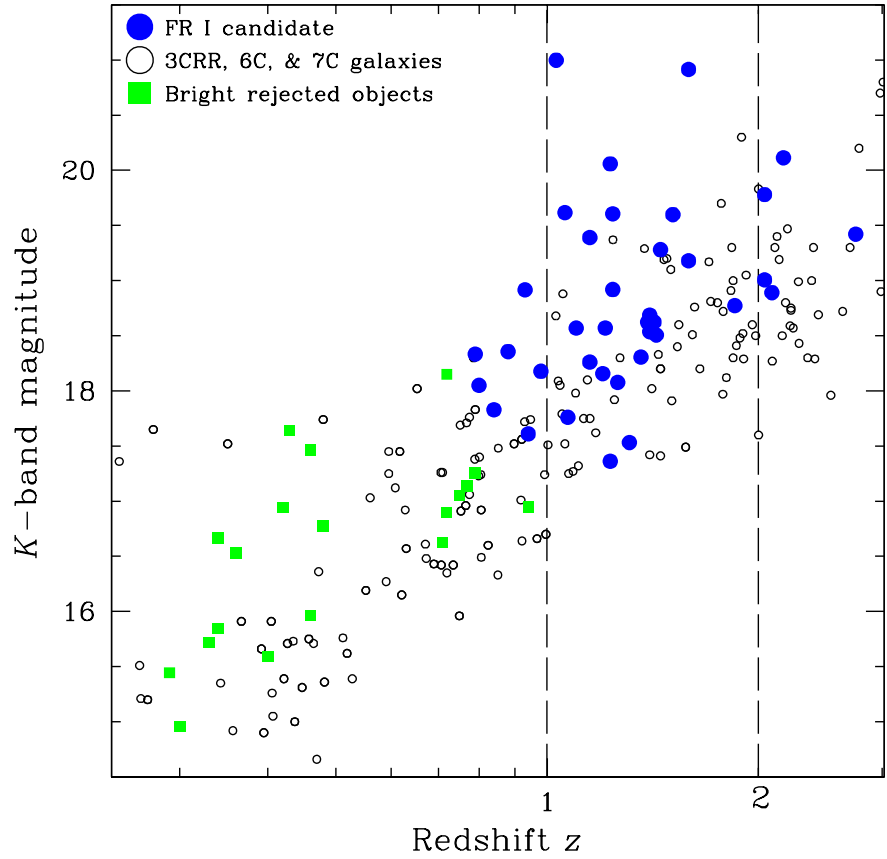}
\caption{The K-z  relation for radio  galaxies. 3C, 6C and  7C sources
from  \citet{willott03}   are  plotted  as  open   circles.  Our  FR~I
candidates are filled circles.  Objects rejected by host galaxy optical
magnitude are plotted as squares.}
\label{fig:kz}
\end{figure}

\subsection{The $K-z$ relation}

Radio  galaxies  are  known  to  obey  the  so-called  $K-z$  relation
\citep{lillylongair82},  a  correlation  between the  infrared  K-band
magnitude    and    redshift,   up    to    at    least   $z\sim    4$
\citep[e.g.][]{jarvis03}. The  origin of  the $K-z$ relation  is still
unclear, and  it is possible that  it is just the  result of selection
effect  \citep{lacy04}.   Our  selection  criteria  make  use  of  the
optical-IR properties  of the hosts of known  powerful radio galaxies,
which  obey the  $K-z$ relation.   However, while  it is  important to
reject ``bright'' optical hosts  in order to eliminate nearby galaxies
from the sample, our selection  criteria do not impose any lower limit
to the  K-band flux of  the hosts.  In  other words, objects  that lie
significantly above  or below the  $K-z$ relation are included  in our
final sample,  if they exist at  all. 

In  Fig.~\ref{fig:kz} we  plot the  $K_s$-band magnitude  of  our FR~I
candidates'  optical  counterparts  vs.   their  photometric  redshift
$z_{\mathrm{phot}}$, as derived from the COSMOS photometric catalog by
\citet{mobasher07}.   The  FR~I   candidates  are  plotted  as  filled
circles.  The  objects that have $m_I  < 21$, are  plotted as squares.
The candidates  morphologically identified as FR~IIs  are not plotted,
as in this paper we only focus on FR~Is.

In addition  to our candidates,  we also plot in  Fig~\ref{fig:kz} (as
empty  circles) the galaxies  from the  3CRR, 6C,  and 7C  catalogs of
radio sources  \citep{willott03}.  Whereas for our  FR~I candidates we
plot  $K_s$-band magnitude  vs.   {\it photometric}  redshift, for  the
\citet{willott03}   data  we   plot  $K$-band   magnitude   vs.   {\it
spectroscopic}  redshift. The brightness  difference between  $K$- and
$K_s$- bands is typically less  than a tenth of a magnitude, therefore
the   minor  differences   that  arise   from   comparing  photometric
measurements in these two bands are not a concern.

As  expected  from our  selection  criteria,  the  objects plotted  as
squares are  relatively nearby galaxies.  With only  a few exceptions,
the FR~I candidates approximately reside in the right redshift bin and
lie on the  $K-z$ relation or slightly above  it.  One outlier (object
70) is  at $z_{\rm  phot} =  2.75$,  which is  unexpected because  our
selection procedure should exclude galaxies at $z \gtrsim 2.5$ (u-band
dropouts),  hence  the  photometric  redshift  for object  70  may  be
incorrect.  A few  objects are $K \sim 2$mag  fainter than the average
of the galaxies of \citet{willott03}.  These might be objects for which the
photometric  redshift is  incorrect.  Alternatively,  they might  be a
population  of radio  galaxies associated  with fainter  hosts.  A few
objects with $K  \sim 2$mag fainter than the bulk  of the radio galaxy
population is in fact observed at low redshifts as well.

In the following section we  describe the FR~I candidates, focusing on
their  radio   morphology,  properties  of  the   host  and  Mpc-scale
environment.

\subsection{Radio Morphology}

Here we  examine the structure  of the 1.4  GHz emission for  those 37
FR~I  candidates  appearing as  filled  circles in  Fig.~\ref{fig:kz}.
Even at  the resolution of the  VLA-COSMOS survey most  of our targets
appear  as compact  radio sources.   Nine of  them show  a discernible
radio  morphology  on  scales   larger  than  $\sim  5\arcsec$,  which
corresponds to a physical scale of $\sim 40$ kpc at $z=1.5$.

We  show the  1.4  GHz COSMOS  VLA  images of  these  nine sources  in
Fig~\ref{extended} (left panels), alongside the $i$-band {\it HST}/ACS
imaging  of their  optical  counterparts (right  panels). The  optical
counterparts are marked with white circles in the HST images.

Even among these  sources with extended radio morphology,  only two of
them (objects 04  and 234) have the radio  morphology of ``bona fide''
FR~Is, in  which the surface brightness  peaks near the  center of the
source, and extended jets are visible.   In objects 02, 18 and 202 the
jet seems to be one-sided and curved, similarly to many ``asymmetric''
nearby FR~I galaxies, as they would appear in shallow observations and
with     poor      spatial     resolution     \citep[e.g.      3C~66B,
3C~129][]{vanbreugel82}.  Objects 18  and 26  have a  radio morphology
similar   to   the    compact   radio   sources   named   ``core-jet''
\citep{conway94},  in which  the visible  jet component  is  almost as
bright  as the radio  core.  The  absence of  the counter-jet  in these
objects might be due to  relativistic beaming effects, if the jet axis
is close to the line-of-sight to the observer.

Object 20  has a very peculiar  morphology. It shows  a bright compact
component,  with a possible  short jet  pointing approximately  to the
N-W.   The ``small-scale'' morphology  is embedded  in a  larger scale
structure  with  lower  surface  brightness, similar  to  an  elongate
``lobe''. Its  peculiar morphology might also  look like a  lobe of an
FR~II, in  which the brightest  region is the  so-called ``hot-spot'',
usually interpreted as the location where the relativistic jet impacts
the  ISM/ICM.  Therefore,  we checked  whether a  ``counter-lobe'' was
present at some  distance along the NW-SE direction.   No radio source
that could be reasonably interpreted as the ``counter-lobe'' is found,
even allowing for the counter-jet  to be slightly bent.  That, and the
clear correspondence of the brightest radio component with a galaxy in
the  HST image, lead  us to  rule out  that this  object could  be the
hotspot of an FR~II.

All  other 28  sources  are slightly  resolved (Fig~\ref{compact})  or
unresolved (Fig~\ref{unres}).  However,  these observations are at 1.4
GHz, which at $z=1.5$ corresponds  to rest-frame 3.5 GHz. The emission
at  that frequency is  dominated by  the central  region of  the radio
source, where young, high energy electrons reside and emit synchrotron
radiation up  to high radio  frequencies, with flat  ($\alpha \lesssim
0.5$,  where $\alpha$  is  defined as  $S_\nu \propto  \nu^{-\alpha}$)
radio spectra.  Extended radio components  such as jets and lobes have
steeper spectra,  thus the emission drops as  frequency increases.  It
may therefore  be that we simply  do not detect  extended, low surface
brightness radio  jets and  lobes for the  majority of  our candidates
given the high rest-frame radio  frequency of the observations and the
intrinsic faintness of the extended regions.

Note also that the size  of the radio sources with ``FR~I-like'' radio
structures is  smaller than the typical  size scales for  FR~Is at low
redshift.   Whereas the  largest  structure we  observe  among our  37
candidates  is  of  order 100  kpc  from  end  to end  (candidate  04,
Fig.~\ref{extended}),  FR~Is  in  the  nearby universe  are  known  to
exhibit larger morphologies, up to a few hundreds of kpc, and in a few
cases            even           Mpc,           \citep[e.g.~B2~1108+27,
NGC~6251][]{perley84}.

Although it is likely that the non-detection of large-scale structures
is a result of the high frequency at which the COSMOS observations are
performed, it is  also possible that our high-$z$  FR~I candidates are
intrinsically small.   In fact, even  the higher power FR~IIs  in this
redshift range appear smaller  than their lower redshift counterparts.
From the work  by e.g.  \citet{kapahi85,gopalkrishna87,kapahi89} it is
known that the projected  linear distance between the hotspots appears
to scale roughly as $(1+z)^\sigma$, where $\sigma$ is between 1 and 2.

Interestingly,   \citet{drake04}   have    found   a   population   of
infrared-bright  radio  sources   that  morphologically  resemble  the
so-called compact  steep spectrum (CSS) sources. CSSs  are believed to
be  young radio  source  that  will eventually  evolve  in the  large,
powerful FR~II  radio galaxies. The sources of  \citet{drake04} are at
least 1 dex less powerful  than normal CSSs, and the derived expansion
velocities of the radio  sources are also significantly smaller.  From
the  analysis of their  overall properties,  these authors  claim that
those mini-radio sources will not  evolve into either FR~IIs or FR~Is,
but will instead lose their radio-loudness and will become radio quiet
FIR-luminous AGNs. Should our  candidates be intrinsically small, they
might  ``fill the  gap''  between  the powerful  CSS  sources and  the
radio-faint  sources of  \citet{drake04}, from  the point  of  view of
their radio properties.  It is  thus possible that most of our targets
are  just  the  progenitors of  the  FR~Is  we  observe in  the  local
universe.

Summarizing, unlike  local FR~Is, the vast majority  of our candidates
show very little  extended morphology in the radio  band.  Clearly, it
is mandatory to investigate whether that is only due to the high radio
frequency at which the observations were made, to the faintness of the
extended  jets and  plumes, or  to the  fact that  our objects  may be
intrinsically small and  possibly young radio sources. The  use of the
EVLA and ALMA will be  necessary in order to achieve sufficient signal
to noise ratio and spatial resolution to study these faint and distant
objects in more details in the radio band.

\subsection{Host Galaxies and environment}

We now discuss the optical sources we have identified as host galaxies
for the candidate FR~I.  Optical images for each of these galaxies are
shown in Figs.~\ref{extended}, \ref{compact} and \ref{unres}. The  photometric
properties of  the hosts, as derived by  \citet{capak07}, are reported
in in Table \ref{tab1}.

\begin{figure}
\plotone{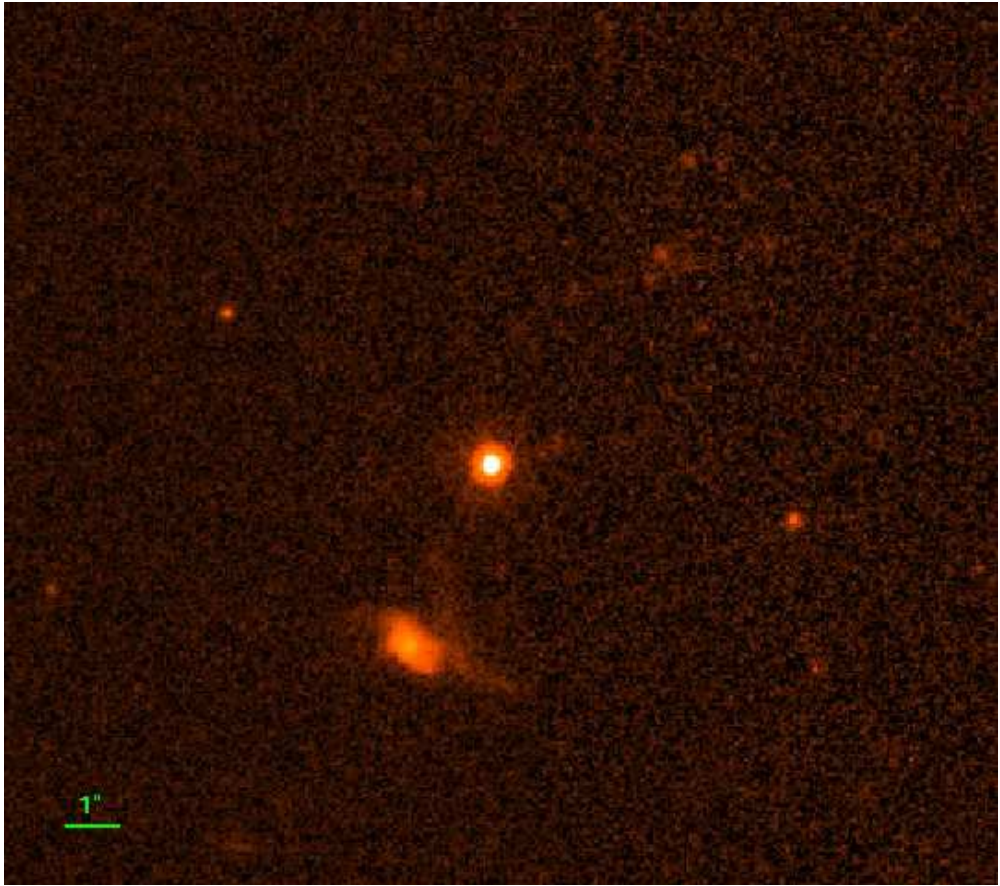}
\caption{HST/ACS F814W  image of target 236.  The optical counterpart
of this object  appears as ``stellar-like''. This might  be an example
of a low power radio galaxy associated with a QSO.}
\label{qso}
\end{figure}

Even if the  images are single orbit HST pointings,  we can attempt to
classify  the  hosts  into  four  different classes,  based  on  their
appearance: (1)  smooth ellipticals, (2) complex, (3)  compact and (4)
unresolved. A more detailed study will be performed when deeper images
are obtained.  Class 1 are  objects of apparent elliptical shape, with
very  little or  no disturbed  morphology;  class 2  are objects  that
appear to  be interacting with close companions  and/or show irregular
morphologies;  class 3  are  barely resolved  galaxies,  too small  to
discern  their  properties;  class  4  includes the  object  that  we
classify as a possible QSO, i.e.  a point source is the only obviously
detected feature (object 236, Fig.~\ref{qso})

From visual  inspection of the images  of the 31  sources with optical
counterparts, 6  of our objects  appear as smooth ellipticals,  10 are
complex,  14 are  compact, and  1 is  unresolved  (stellar-like).  The
resolution of the images of the 6 host galaxies that are only detected
in   the  IR   does  not   allow  us   to  derive   any  morphological
classification.  It is worth noting  the high fraction of objects with
complex optical morphologies among  our FR~I candidates.  This appears
to be at odds with low redshift FR~Is, the large majority of which are
hosted  by undisturbed  ellipticals or  cD  galaxies \citep{zirbel96}.
However, the larger fraction of complex morphologies in our sample may
simply  reflect the  different stage  of evolution  of the  host, that
might still be in a very active merging phase at $1 < z < 2$.

Not much can be said about  the compact galaxies. Because of the short
exposure times  of the HST data, we  might just be seeing  the core of
the  galaxies, while  we are  missing  the external  regions of  lower
surface  brightness.   Alternatively,  these  might  be  intrinsically
smaller  objects, which  would  contrast with  local  FR~I hosts  that
appear to be invariably associated with giant ellipticals.

The  presence   of  one  stellar-like  optical   counterpart  is  also
intriguing.  We  interpret this object as a  compact nucleus (possibly
the  AGN)  outshining  the  host  galaxy.   This  corresponds  to  the
morphology observed  in QSOs, but scaled  down in luminosity  by a few
orders  of  magnitude.   
However,  it has  been  established  that at  low  redshifts no  FR~Is
belonging to  the 3CR \citep{spinrad} or  B2 \citep{b2catalog} samples
are associated with host galaxies of this kind. Most importantly, from
the point of view of the physics of their active nucleus, none of them
appear to  show broad lines  in the optical  spectrum\footnote{This is
true with the only exception  of the peculiar galaxy 3C~120, an object
with  FR~I radio  morphology  associated  with an  S0  host showing  a
spiral-like structure. However, 3C~120 is formally not part of the 3CR
catalog of  \citet{spinrad}.}.  Recently, \citet{zamfir08}  have found
no FR~I-QSO in a  large sample of SDSS-FIRST/NVSS quasars, reinforcing
the idea that FR~I-QSOs are extremely rare in the local universe.

Clearly, this ``nucleated'' galaxy need further investigation aimed at
determining  its nature  as a  quasar through  the detection  of broad
permitted lines in its  rest-frame optical spectrum.  However, even if
the unresolved  object was spectroscopically confirmed as  a QSOs, the
fraction of FR~I  quasars in our sample (1/37)  would be significantly
lower  than  the  fraction of  FR~II  quasars  in  the same  range  of
redshift, which is  around 40\% \citep[e.g.][]{willott00}.  A possible
scenario is that  the smaller fraction of FR~I-QSO  as compared to the
fraction  of FR~II-QSO  simply  results from  the  dependence of  such
fraction  on   luminosity  \citep{willott00}.   This   may  reflect  a
reduction  of  the  opening   angle  of  the  ``obscuring  torus''  as
luminosity  decreases   (the  so  called   ``receding-torus''  model).
Alternatively,  most high-z FR~Is  may intrinsically  lack significant
broad emission  line region  and thermal disk  emission as is  for the
FR~Is  in low-z  samples  \citep[e.g.][]{pap1}.  These  issues can  be
explored with deep imaging to  determine the nature of the hosts using
HST and with spectroscopy using  an 8m-class telescope to determine the
presence or absence of any strong broad emission lines.

Although our  sample cannot be considered  statistically complete, the
selection  criteria  are not  biased  against  the  presence of  QSOs.
Instead, the selection  based on the radio flux  at 1.4MHz is somewhat
biased  in favor of  core-dominated, relativistically  beamed objects,
and  none of  the objects  that  were rejected  because their  optical
i-band magnitude exceeds our  selection limit were point sources.  The
existence  of  a large  number  of  FR~I-QSOs at  intermediate-to-high
redshifts  has been noted  by \citet{heywood07}.   This would  imply a
strong  evolution in the  physical properties  of radio  galaxies with
FR~I radio morphology, since FR~I-QSOs are known to essentially absent
at low redshifts.  However,  besides the different selection criteria,
the objects of \citet{heywood07}  are mostly high power sources, while
here we focus on radio galaxies  of the same power as low-z FR~Is.  It
is thus possible that at  high-z, the FR~I break shifts towards higher
radio powers.

An in-depth  analysis of the properties  of the objects  in our sample
and  the implications  for  the  AGN unification  scheme  will be  the
subject   of  future   work.   However,   the   limited  morphological
information  we  have  at  this  stage seems  to  show  that  FR~I-QSO
represent  a  tiny  fraction  of  the low-power  FR~I  radio  galaxies
population  at  $1 <  z  <  2$.

\subsection{FR~I candidates as tracers of high-$z$ galaxy clusters?}
\label{clusters}

\begin{figure}
\epsscale{0.8}
\plotone{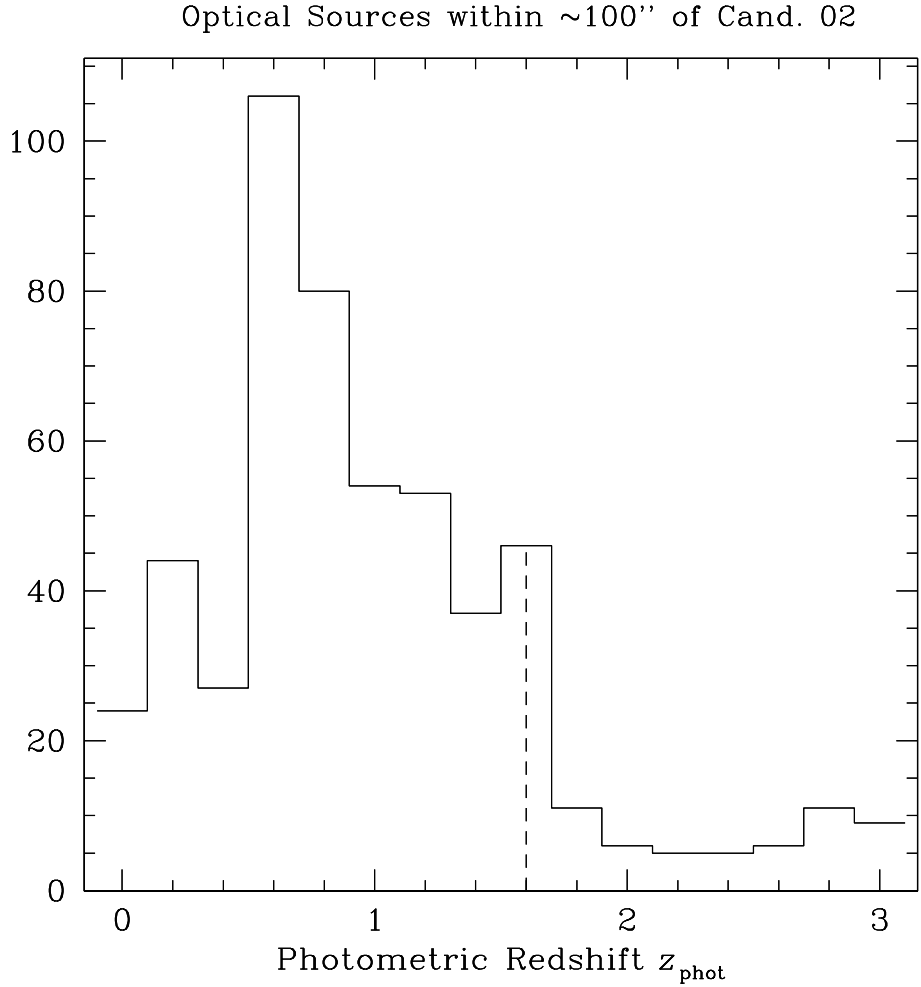}
\caption{Photometric redshifts distribution  of optical sources in the
COSMOS catalog  within 100$^{\arcsec}$ of candidate  02.  The vertical
dashed line  marks the photometric  redshift of the candidate  FR~I. A
peak in the redshift distribution corresponding to the redshift of our
source  is evident,  and  may be  interpreted  as the  presence of  an
overdensity of galaxies at the redshift of the FR~I candidate.}
\label{cluster02}
\end{figure}

\begin{figure*}
\epsscale{1.1}
\plotone{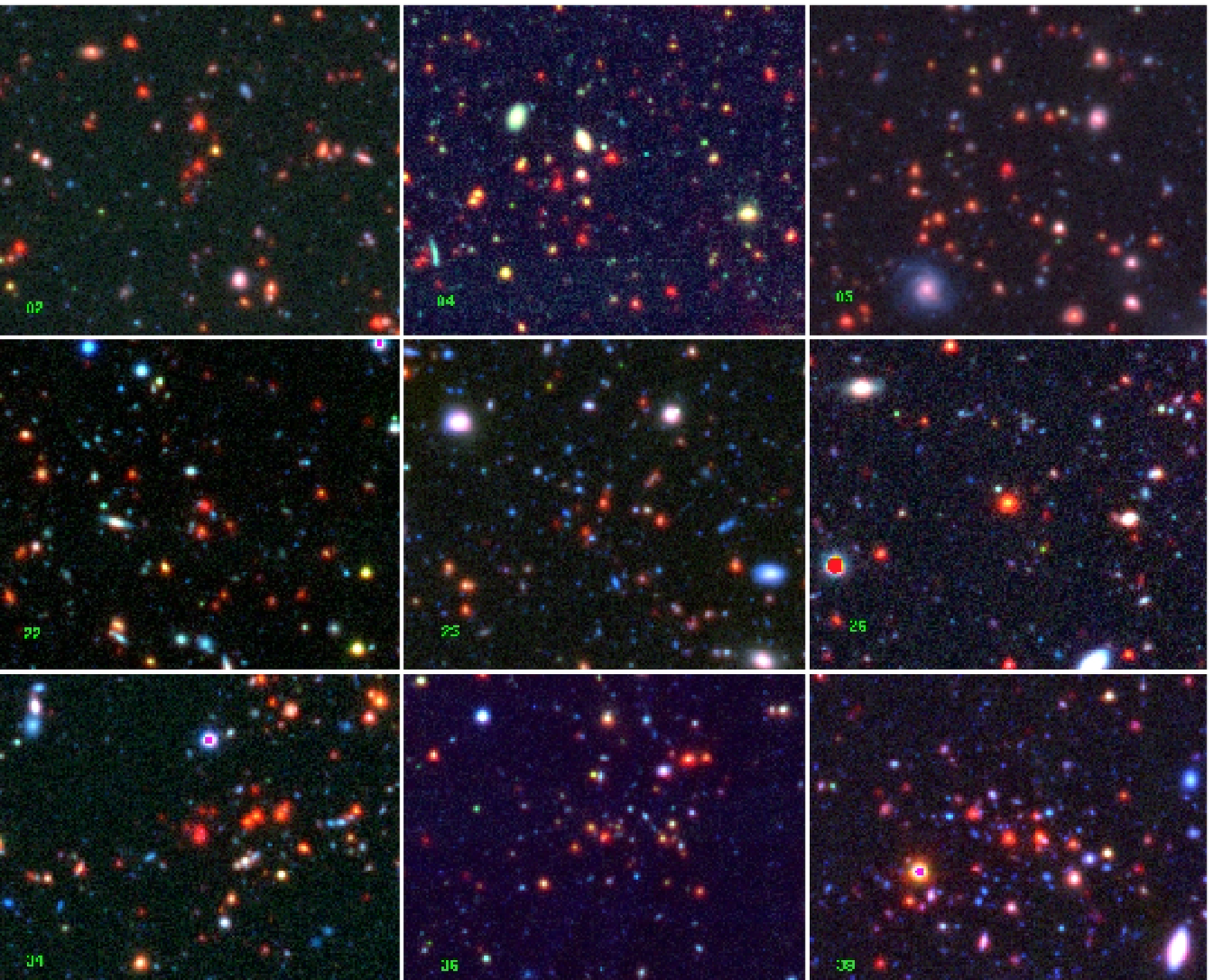}
\caption{RGB images of nine cluster candidates found around our high-z
FR~I candidates. The ``color'' images are obtained using {\it Spitzer}
data at  3.6$\mu$m for the  R channel, z-band  for the G  channel, and
V-band for the  B channel. The projected scale of each image is $\sim
110^{\arcsec} \times 90^{\arcsec}$}
\label{cluster1}
\end{figure*}

One of the  motivations for our high-z FR~I radio  galaxy search is to
locate  high-z clusters  of galaxies.   In the  local  universe, $\sim
70\%$ of  the entire  population of FR~Is  is associated  with cD-like
galaxies  \citep{zirbel96},  and  almost  all low-z  FR~Is  reside  in
clusters  of  various   richness  \citep{zirbel97}.   Only  $\sim  15$
clusters are  known to exist between  $1<z<2$, thus a  large sample of
FR~Is may easily double the number of clusters in that redshift range,
assuming that  the environment of FR~Is does  not significantly change
with redshift.  Although we defer  a systematic search for clusters to
a forthcoming paper, in this  section we qualitatively explore the Mpc
environments of our candidate FR~I,  in order to probe the possibility
that their  host galaxies reside  in clusters, similarly to  their low
$z$ counterparts.

In order  to look  for cluster candidates,  we followed  two different
methods.  Firstly, we searched  the COSMOS catalog for the photometric
$z$ of objects that are located  inside a region of projected radius 1
Mpc from  the FR~I candidate.  Only in  the case of object  02 we find
qualitative  evidence  for  an  over-density of  galaxies  around  the
redshift  of our  FR~I candidate,  which  we interpret  as a  possible
presence of a group or a  cluster. In Fig. \ref{cluster02} we show the
photometric   redshifts  distribution   of   optical  sources   within
100$^{\arcsec}$  (corresponding to  a  projected radius  of 855kpc  at
$z=1.6$)  of  candidate  02.   The  vertical  dashed  line  marks  the
photometric redshift  of the  FR~I candidate. A  peak in  the redshift
distribution corresponding  to the redshift of our  source is evident.
However, the low detection rate  obtained with the above method is not
surprising. Our FR~I candidates are quite possibly among the brightest
cluster members, while  an $L^*$ galaxy at $z\sim  1.5$ is expected to
have $m_I \sim  24$.  Thus, the COSMOS optical images  are in fact not
deep  enough to  detect a  significant number  of cluster  galaxies at
redshifts $z>1$.  And even when they are detected in the images, their
photometric $z$ can be highly uncertain because of the large errors on
photometry of such faint objects.

Another, more effective, method we explored to find cluster candidates
using  the data  from COSMOS,  is to  look for  extremely  red objects
around  our FR~I  candidates. Although  there is  still some  level of
degeneracy between  objects that are intrinsically  red and redshifted
objects, this method seems to  lead to more promising results, even at
the qualitative level presented here.  In Fig.  \ref{cluster1} we show
nine cluster candidates found with the latter method.  We produced RGB
``color'' images using {\it Spitzer Space Telescope} data at 3.6$\mu$m
for the  R channel,  z-band for the  G channel,  and V-band for  the B
channel.  We are  setting  up observational  programs  to measure  the
redshift of  the brightest cluster members and  we will systematically
study cluster candidates in a forthcoming paper.

In order to achieve more  detailed information about the morphology of
the  host   galaxies,  their  possible   interactions  with  immediate
neighbors,  and  the  properties  of  the  cluster  environment,  deep
high-resolution  optical  and  near-IR  images and  possibly  slitless
spectroscopy should  be obtained as soon  as the WFC3  is installed on
HST.  Only the future  generation of high-sensitivity X-ray telescopes
(e.g. IXO) will allow us to  study in detail the properties of any hot
virialized gas in the cluster environment of our sources.  However, in
a forthcoming paper we will study the {\it Chandra}/COSMOS data to try
and detect such a hot gas using stacking techniques.

\section{Summary \& Conclusions}
\label{conclusions}

We have  outlined our search for  FR~I radio galaxy  candidates in the
COSMOS field at redshift $1<z<2$.  Previously, no low power FR~I radio
galaxies were  known in  this redshift bin,  besides one  (or possibly
two)  candidate in  the HDF  North \citep{snellenbest}.   Flux limited
samples  are not  suitable  for finding  low-power  radio galaxies  at
high-z,   because  of   the  tight   redshift-luminosity  correlation.
Therefore,  we  used  a  4-steps multi-wavelength  selection  process,
starting  from radio flux,  and using  radio morphologies  and optical
magnitudes to further  constrain the sample selection.  At  the end of
the selection process we are left with a sample of 37 objects.

The photometric redshift of the bulk of our FR~I candidates are in the
expected  range  $1  < z_{\rm  phot}  <  2$.   The redshifts  must  be
confirmed with  spectroscopic observations possibly using  at least an
8m-class   telescope,  future   larger   instruments  or   space-based
observations  to take  advantage  of the  lower  background.  In  most
cases, the radio images show objects with compact morphologies.  These
might be intrinsically young sources, that will eventually evolve into
the   giant   FR~I  radio   galaxies   observed   at  low   redshifts.
Alternatively, the  extended emission is  not detected because  of the
rest-frame  high  frequency  at  which the  observations  were  taken.
Further investigation  is needed  to address this  issue.  We  plan on
obtaining  low-frequency  radio observations  to  detect any  extended
radio  emission from  ``older''  electrons, combined  with deep,  high
resolution data  at a higher wavelength  ($\sim 8$ GHz)  to derive the
spectral index of the source and to study the morphology.

The short  one-orbit i-band  HST observations are  not suitable  for a
detailed morphological study of  the host galaxies.  However, the data
show  variegated  morphologies, ranging  from  smooth ellipticals  to
complex interacting systems.   A few of them appear  to be compact and
one  is stellar-like.   This object  might belong  to a  population of
FR~I-QSOs  that  is  basically  not  present in  the  local  universe.
However, the fraction of low-power  FR~I-QSOs in our sample appears to
be significantly lower than the  overall fraction of FR~II-QSOs in the
same redshift bin. Optical-IR spectroscopy of the sources is needed to
assess the nature of this candidate QSO.

Although the images from the  COSMOS survey are not suitable to detect
a large fraction of cluster galaxies at $z>1$, the environment of some
of  our  FR~I  candidates  shows   evidence  for  the  presence  of  a
cluster. This  is apparent when the  IR images from  the {\it Spitzer}
space telescope are used in  combination with the optical ground based
data, resulting in a significant number of red objects surrounding the
host galaxies of our FR~I candidates.

The  search for  high-z FR~I  candidates  we presented  in this  paper
constitutes a pilot study for objects to observed with high-resolution
and  high-sensitivity  future instruments.   The  EVLA  and ALMA  will
provide us  with crucial  information on the  radio morphology  of our
sources, they will help us to understand whether or not the objects we
discovered are intrinsically small, and if they are ``progenitors'' of
the local FR~I population.  When WFC3 is installed on the HST, it will
be possible  to study  in greater details  the properties of  the host
galaxies  and the  cluster environment  in  the optical  and IR,  with
important bearings for our knowledge of the origin of the most massive
galaxies  and  galaxy  clusters.    Clearly,  these  studies  will  be
complemented and further expanded when JWST will be available, as well
as when  future generation  high-sensitivity X-ray satellites  will be
launched.

\acknowledgments We are indebted with the referee for her/his comments
that greatly  improved the  paper.  We are  grateful to  Piero Rosati,
Stefi  Baum and Meg  Urry for  helpful discussions.   GRT acknowledges
support  from HST-GO-10882.01-A.   We  acknowledge the  effort of  the
entire  COSMOS  team without  which  this  work  would not  have  been
possible.     More   information   on    COSMOS   is    available   at
\texttt{http://cosmos.astro.caltech.edu}.  The  {\it Very Large Array}
is a facility of the National Radio Astronomy Observatory, operated by
Associated  Universities, Inc., under  cooperative agreement  with the
National Science Foundation.  This research  has made use of the NASA/
IPAC Infrared Science Archive, which is operated by the Jet Propulsion
Laboratory,  California Institute of  Technology, under  contract with
the National Aeronautics and Space Administration.



\end{document}